\documentclass[aps,pre,twocolumn,showpacs,superscriptaddress]{revtex4-2}
\usepackage{graphicx}  
\usepackage{dcolumn}   
\usepackage{bm}        
\usepackage{dsfont}    
\usepackage{amssymb}   
\usepackage{amsmath}
\usepackage{color}
\usepackage{siunitx}
\usepackage{physics}
\usepackage{hyperref}
\usepackage{xcolor}

\usepackage[utf8]{inputenc}

\newcommand{\chg}[1]{#1}

\definecolor{hawaii}{HTML}{00452a} 
\newcommand{\chgtwo}[1]{#1}

\DeclareRobustCommand{\okina}{%
	\raisebox{\dimexpr\fontcharht\font`A-\height}{%
		\scalebox{0.8}{`}%
	}%
}
\newcommand{\hawaii}{Hawai\okina i}

\begin{document}
	\title{Data-Induced Interactions of Sparse Sensors Using Statistical Physics}
	\author{Andrei A.~Klishin}
        \email{aklishin@hawaii.edu}
        \affiliation{Department of Mechanical Engineering, University of \hawaii~at M\=anoa, Honolulu, HI 96814, USA}
	\affiliation{AI Institute in Dynamic Systems, University of Washington, Seattle, WA 98195, USA}
	\affiliation{Department of Mechanical Engineering, University of Washington, Seattle, WA 98195, USA}
	\author{J.~Nathan Kutz}
	\affiliation{AI Institute in Dynamic Systems, University of Washington, Seattle, WA 98195, USA}
	\affiliation{Departments of Applied Mathematics and Electrical and Computer Engineering, University of Washington, Seattle, WA 98195, USA}
	\author{Krithika Manohar}
        \email{kmanohar@uw.edu}
	\affiliation{AI Institute in Dynamic Systems, University of Washington, Seattle, WA 98195, USA}
	\affiliation{Department of Mechanical Engineering, University of Washington, Seattle, WA 98195, USA}

	\date{\today}
	
	\begin{abstract}
		Large-dimensional empirical data in science and engineering frequently have a low-rank structure and can be represented as a combination of just a few eigenmodes. Because of this structure, we can use just a few spatially localized sensor measurements to reconstruct the full state of a complex system. The quality of this reconstruction, especially in the presence of sensor noise, depends significantly on the spatial configuration of the sensors. Multiple algorithms based on gappy interpolation and QR factorization have been proposed to optimize sensor placement. Here, instead of an algorithm that outputs a single ``optimal'' sensor configuration, we take a statistical mechanics view to compute the full landscape of sensor interactions induced by the training data. \chgtwo{The two key advances of this paper are the recasting of the sensor placement landscape in an Ising model form and a regularized reconstruction that significantly decreases reconstruction error for few sensors. In addition, we provide first uncertainty quantification of the sparse sensing reconstruction and open questions about the shape of reconstruction risk curve.} Mapping out these data-induced sensor interactions allows combining them with external selection criteria and anticipating sensor replacement impacts.
        
	\end{abstract}
	
	\maketitle

\section{Introduction}
Many natural and engineered systems can take a variety of high-dimensional states, with the amount of data growing rapidly with the number of observed snapshots and increasing snapshot resolution. At the same time, the amount of information in this data usually grows much slower, often logarithmically \cite{quinn2022information, udell2019big, gavish2014optimal}. In this situation, any system state can be closely approximated by a combination of just a few basis vectors, enabling algorithms from lossy image compression to Dynamic Mode Decomposition \cite{kutz2016dynamic, lewis1992image}. While the optimal basis can be learned from historical data or high-fidelity simulations, states cannot be measured in that basis directly and often can only be accessed by spatially-localized sensors.

Reconstruction of full states from localized sensor measurements has a long history under the umbrella term of \emph{compressed sensing}, where the sampling points (sensor locations) are chosen randomly and the state is reconstructed as a sparse combination of universal basis vectors \cite{donoho2006compressed, ganguli2010statistical, krzakala2012statistical}. More recently, driven by advances in gappy and reduced-order \chg{Partial Differential Equation (PDE)} methods \cite{everson1995karhunen, barrault2004empirical, chaturantabut2009discrete}, \emph{sparse sensing} algorithms have been developed to take advantage of the available training data to reduce the number of sensors required for given reconstruction quality \cite{manohar2018data}. The general sparse sensing problem is usually set up as follows: given the training data matrix $\mathbf{X}$ consisting of $N$ snapshots of an $n$-dimensional state, one needs to reconstruct an unknown state $\mathbf{x}\in \mathbb{R}^n$ sampled from the same distribution as the data by using only the noisy measurements of a few components of the state $\mathbf{y}\in \mathbb{R}^p, p\ll n$.

While any combination of sensors of appropriate rank can be used to compute the maximal likelihood state reconstruction, the reconstruction robustness to sensor noise may vary by orders of magnitude, leading to the problem of \emph{sensor placement}. Each sensor configuration can be assigned a cost function value that can be approximately maximized with efficient greedy heuristics based on optimal experiment design, information theory metrics, Gibbs sampling, or matrix QR pivoting \cite{deaguiar1995doptimal, krause2008near, sun2020learning, peherstorfer2020stability, manohar2018data}. While these methods return a sensor configuration provably close to the true optimum due to the submodularity property \cite{nemhauser1978analysis, krause2011submodularity, otto2022inadequacy}, they do not inform why a particular configuration should be chosen, how to best modify it if sensor budget changes, and what would be the impact of a sensor malfunction on the reconstruction quality.

\chg{Reconstruction of the full state from few measurements is a type of \emph{inverse problem}, which is inherently ill-conditioned and needs to be regularized \cite{kaipio2006statistical}. Here we regularize the reconstruction by framing it as Bayesian inference combining a prior distribution with sensor measurement likelihood into a posterior distribution over the reconstructed states. While this approach has not been used before in sparse sensing literature \cite{manohar2018data}, it is common in Optimal Experimental Design (OED). Bayesian OED proposes some functional of the posterior distribution, such as expected information gain, as the utility function that can be optimized over experimental designs \cite{chaloner1995bayesian}. However, most of the effort is focused on \emph{evaluating} the utility for generic design variables as opposed to offering practical methods of \emph{optimizing} it \cite{chaloner1995bayesian, ryan2016review, rainforth2024modern}. Some Bayesian OED is framed in terms of sensor selection but relies on expensive methods such as Monte Carlo and $L_1$ sparsity penalty \cite{alexanderian2021optimal}.}

\chg{Whereas Bayesian OED is focused on searching for a single ``optimal'' sensor configuration, in this paper}
we take a statistical mechanics perspective to \chg{decompose the design objective and} study the entire landscape of sensor interactions, \chg{similar} to saliency maps in machine vision \cite{simonyan2013deep}. We show that the sensor interactions can be interpreted in terms of 1-body, 2-body, and higher order Hamiltonian terms  computed directly from the training data. Understanding the part of the landscape induced by data directly inspires a greedy sensor placement algorithm, allows incorporating landscapes driven by external cost factors, and anticipates the impacts of sensor replacement needs. The energy landscape analysis can be combined with other recent advances in sensor placement studies.

\chgtwo{The two key advances of this paper are the sensor placement landscape decomposition in an Ising model form and the regularized reconstruction formula. We also derive results for reconstruction uncertainty heatmap and note the appearance of the double descent phenomenon in reconstruction risk curves.} The rest of the paper is organized as follows. In Section~\ref{sec:background} we review the key mathematical formalism of low-dimensional representations and sparse sensing of signals. In Section~\ref{sec:methodology} we derive the Bayesian version of state reconstruction and map the cost function of sensor sets to the Ising model. In Section~\ref{sec:results} we benchmark the sensor placement methods and reconstructions on several empirical and synthetic datasets. In Section~\ref{sec:discussion} we discuss the connections of our method with other results across sparse sensing and machine learning and propose extensions for future work.

\section{Background}
\label{sec:background}
The representation of system states as a combination of basis vectors is crucial for state reconstruction from sparse measurements. We distinguish between two types of bases: \chg{universal and tailored}. Universal basis vectors can approximate any signal of sufficient regularity. However, natural signals are often highly compressible and \emph{sparse}, with only a few nonzero basis coefficients.  Compressed sensing leverages this inherent compressibility by retrieving the sparsest nonzero \chg{coefficients}.
Although compressed sensing provides probabilistic recovery guarantees,\chg{\cite{ganguli2010statistical}} it requires more measurements than the actual signal sparsity. Additionally, ensuring that these measurements are randomized or uncorrelated with the basis vectors poses practical challenges.

When information is available about a specific type of signal, using a tailored basis representation becomes advantageous \chg{\cite{manohar2018data}}. Most science and engineering applications operate on highly structured signals governed by controlled physical processes. Such signals are typically low-rank in a tailored basis specifically designed for them, which can be learned from data. Unlike universal bases, where the activated basis functions are not known \emph{a priori}, a tailored basis contains precisely the few basis functions relevant to the data. This reduces the signal representation to a small number of tailored basis coefficients, or coordinates. As a result, operations that are computationally intractable in the ambient data dimension become highly efficient in these new reduced coordinates. 

\chg{A high-dimensional state can be closely approximated in the tailored basis obtained by the Proper Orthogonal Decomposition (POD) of the training data set:}
\begin{align}
	\mathbf{X}=&\mathbf{\mathbf{\Psi} \Sigma V}^T \approx \mathbf{\mathbf{\Psi}}_r \mathbf{\Sigma}_r \mathbf{V}_r^T,\\
    \mathbf{x}\approx&\mathbf{\Psi}_r\mathbf{a}= \sum\limits_{k=1}^{r} \mathbf{\psi}_k \mathbf{a}_k
\end{align}
where we dropped all singular values beyond the first $r$ per the optimal truncation prescription \cite{gavish2014optimal}.
Reduced-order models (ROMs) exploit this dimensionality reduction to project high-dimensional systems of \chg{Ordinary Differential Equations (ODEs)} 
down to a system of $r$ ODEs, enabling highly efficient time-stepping, parameter estimation, and optimal control. 

In the sparse sensing setup, the tailored basis coefficients $\mathbf{a}$ are not known \emph{a priori} and cannot be measured directly. Instead of measurements that are sparse in the POD mode space (since $r\ll n,N$), we have access to sensor measurements that are sparse in the original space (since $p\ll n$):
\begin{align}
	\mathbf{y}=\mathbf{Cx}=\mathbf{C\mathbf{\Psi}}_r \mathbf{a}=\mathbf{\mathbf{\Theta}} \mathbf{a},
\end{align}
where $\mathbf{C}: p\times n$ is the sensor selector matrix with the entry 1 for the location that each sensor measures, and 0 otherwise. 

In the setting $p=r$, discrete empirical interpolation methods (DEIMs)~\cite{chaturantabut2009discrete} equate  sensors to optimized interpolation points in the POD basis, which are represented by the point measurement operator $\mathbf{C}$. DEIMs were developed to approximate nonlinear terms in ROMs, with the basis coefficients given by the exact interpolation of optimized sensor measurements
\begin{align}
\hat{\mathbf{x}} = \mathbf{\mathbf{\Psi}}_r (\mathbf{C\mathbf{\Psi}}_r)^{-1} \mathbf{y}.
\end{align}

When there are more measurements than coefficients, the solution might not pass through the measurements directly and is instead given by regression:
\begin{align}
\hat{\mathbf{x}} = \mathbf{\mathbf{\Psi}}_r (\mathbf{C\mathbf{\Psi}}_r)^{\dagger} \mathbf{y},
\end{align}
where $\dagger$ stands for the Moore-Penrose pseudoinverse of a rectangular matrix. This technique is more generally known as gappy POD~\cite{everson1995karhunen}. While optimized measurements have been devised for this more general setting, such as ODEIM~\cite{peherstorfer2020stability}, DEIM sensing strategies aim to control the condition number of the matrix inversion 
$(\mathbf{C\mathbf{\Psi}}_r)^{-1}$ or $(\mathbf{C\mathbf{\Psi}}_r)^\dagger$, and do not explicitly consider sensor noise. Optimized sensing when 
$p<r$ remains an open question, which is important in applications where sensor measurements are much sparser and heavily constrained. Both concerns are addressed by recasting to a Bayesian estimation method, described \chg{in the next section}.

\chg{Even if the \emph{number} of sensors $p$ is kept constant, the quality of reconstruction depends significantly on \emph{which} sensors are selected. While gradient-based approaches and convex optimization result in fairly good sensor sets, they have an aggressive computational time scaling of $\order{n^3}$ \cite{joshi2008sensor, chen2011h2}. More recent approaches exploited greedy sensor placement based on the QR factorization of the $\mathbf{\Psi}_r$ matrix, resulting in comparable sensor set quality with only $\order{nr^2}$ operations required to place $r$ sensors \cite{manohar2018data, manohar2021optimal}.}

\begin{table}[t]
    \centering
    \begin{tabular}{|ll|}
    \hline
    \multicolumn{2}{|c|}{\Large Scalars} \\
    $n$     & full state dimension \\
    $N$     & number of training snapshots \\
    $r$     & number of modes used for reconstruction \\
    $p$     & number of sensors \\
    $\eta$  & standard deviation of noise \\
    $\sigma_{prior}$    & standard deviation of isotropic prior \\
    \chgtwo{$\sigma_{scale}$}   & \chgtwo{natural scale of a dataset} \\
    \multicolumn{2}{|c|}{\Large Vectors} \\
    $\mathbf{x}\in\mathbb{R}^n$    & true full state \\
    $\hat{\mathbf{x}}\in\mathbb{R}^n$  & reconstructed full state \\
    $\mathbf{\psi}_k$ & $k$th vector of the POD basis \\
    $\mathbf{y}\in\mathbb{R}^p$    & sensor measurements \\
    $\mathbf{a}\in\mathbb{R}^r$    & true latent state \\
    $\hat{\mathbf{a}}\in\mathbb{R}^r$  & reconstructed latent state \\
    $\mathbf{g}_i\in\mathbb{R}^r$  & sensing vector at location $i$ \\
    $\mathbf{\sigma}\in\mathbb{R}^n$   & reconstruction uncertainty heatmap \\
    $\gamma\in\mathbb{N}^p$ & set of sensors \\
    \multicolumn{2}{|c|}{\Large Matrices} \\
    $\mathbf{X}\in\mathbb{R}^{n\times N}$    & training data matrix \\
    $\mathbf{\Psi}\in\mathbb{R}^{n\times n}$ & POD basis \\
    $\mathbf{\Psi}_r\in\mathbb{R}^{n\times r}$   & truncated POD basis \\
    $\mathbf{\Sigma}\in\mathbb{R}^{N\times N}$   & POD singular values \\
    $\mathbf{\Sigma}_r\in\mathbb{R}^{r\times r}$ & truncated POD singular values \\
    $\mathbf{V}\in\mathbb{R}^{N\times N}$    & POD right singular vectors \\
    $\mathbf{V}_r\in\mathbb{R}^{r\times N}$  & truncated POD right singular vectors \\
    $\mathbf{A}\in\mathbb{R}^{r\times r}$    & matrix inverted in reconstruction \\
    $\mathbf{B}\in\mathbb{R}^{n\times p}$    & uncertainty heatmap matrix \\
    $\mathbf{S}\in\mathbb{R}^{r\times r}$    & reconstruction prior covariance matrix \\
    $\mathbf{C}\in\mathbb{R}^{p\times n}$    & sensor selection matrix \\
    $\mathbf{\Theta}\in\mathbb{R}^{p\times r}$   & $\mathbf{C\Psi}_r$ product \\
    $\mathbf{G}\in\mathbb{R}^{p\times r}$    & matrix of selected sensing vectors \\
    $\mathbf{D}\in\mathbb{R}^{p\times p}$    & diagonal part of $\mathbf{GG}^T$ \\
    $\mathbf{R}\in\mathbb{R}^{p\times p}$    & off-diagonal part of $\mathbf{GG}^T$ \\
    \hline
    \end{tabular}
    \caption{Notation table.}
    \label{tab:notation}
\end{table}

\section{Methodology}
\label{sec:methodology}

In this section we derive (i) Bayesian estimation of $\hat{\mathbf{a}}$ from \emph{any} set of sensor locations and vector of sensor measurements $\mathbf{y}$, and (ii) the landscape of sensor placement that selects for minimal reconstruction error induced by sensor noise. The mathematical notation is explained in Table~\ref{tab:notation}.


\subsection{Bayesian inference}
Here we situate state reconstruction as a problem of Bayesian inference from noisy sensor data. We place $p\ll n$ sensors each measuring the state in one location. We denote the set of sensors as $\gamma$, the set of sensor location indices. Given a full $n$-dimensional state vector $\mathbf{x}$, the noiseless sensor output is a much shorter state vector, sometimes called \chg{gappy} \cite{everson1995karhunen}:
\begin{align}
	\mathbf{y}=\mathbf{C}\mathbf{x}=\mathbf{C\mathbf{\Psi}}_r \mathbf{a}=\mathbf{\mathbf{\Theta}} \mathbf{a},
\end{align}
where $\mathbf{C}: p\times n$ is the sensor selector matrix with the entry 1 for the location that each sensor measures, and 0 otherwise. The matrix $\mathbf{\mathbf{\Theta}}: p\times r$ combines the sensor selection with the low-rank representation of the state.

We further assume that each sensor measures the state with a Gaussian noise of magnitude $\eta$. Thus, given a true state $\mathbf{a}$, the likelihood of sensor readings is given by:
\begin{align}
	p(\mathbf{y}|\mathbf{a})\propto \exp(-\frac{(\mathbf{y}-\mathbf{\mathbf{\Theta}} \mathbf{a})^2}{2\eta^2}),
\end{align}
where we omit the distribution normalization.

For Bayesian inference, we invert the distribution using the Bayes rule:
\begin{align}
	p(\mathbf{a}|\mathbf{y})=\frac{p(\mathbf{y}|\mathbf{a})p(\mathbf{a})}{p(\mathbf{y})},
\end{align}
where $p(\mathbf{a})$ is a \emph{prior} distribution and $p(\mathbf{y})$ is a normalization. The procedure of state estimation consists of computing the Maximum A Posteriori (MAP) estimate, which is typically done on log-\chg{posterior}:
\begin{align}
	\hat{\mathbf{a}}=\arg\max\limits_{\mathbf{a}}\left( \ln p(\mathbf{y}|\mathbf{a}) + \ln p(\mathbf{a}) \right),
\end{align}
where the normalization $p(\mathbf{y})$ was omitted as it doesn't depend on the inferred state $\mathbf{a}$. The solution of this argmax problem requires knowing the functional form of the prior that we discuss below.

\subsection{Constructing the prior}
In order to exploit the prior information of the data, we need to assume a functional form of the prior distribution over the coefficients $\mathbf{a}$. A simple form of this assumption is to select a Gaussian prior of the form:
\begin{align}
	p_\textsf{Gauss}(\mathbf{a})\propto \exp(-\frac{\mathbf{a}^T \mathbf{S}^{-2} \mathbf{a}}{2}),
	\label{eqn:priorgauss}
\end{align}
which posits that the system states are drawn from an anisotropic Gaussian cloud where the variances along each orthogonal direction are given by the elements of a diagonal matrix $\textbf{S}$. We consider two choices for the prior: a scaled identity matrix $\textbf{S}=\sigma_{prior} \mathbf{I}_r$ for isotropic variance along all dimensions, and the matrix of \chgtwo{normalized} singular values of the training data \chgtwo{$\mathbf{S}=\mathbf{\Sigma}_r/\sqrt{N-1}$} for hierarchically decreasing variance of higher modes \chgtwo{following Ref.~\cite{kakasenko2025bridging}}. Since in both cases all $r$ elements are positive, the $\mathbf{S}$ matrix is invertible and thus the prior is normalizable. In this case the prior functions as a regularizer of state reconstruction. More complex prior distributions can be constructed for training data situated on curved manifolds \cite{otto2022inadequacy}.

\subsection{Gaussian prior inference}
For the Gaussian functional form of the prior (Eqn.~\ref{eqn:priorgauss}), we explicitly write out the \chg{log-posterior} as follows:
\begin{align}
	\ln p(\mathbf{a}|\mathbf{y})=-\frac{1}{2\eta^2}\left( \mathbf{y}-\mathbf{\mathbf{\Theta}} \mathbf{a} \right)^T \left( \mathbf{y}-\mathbf{\mathbf{\Theta}} \mathbf{a} \right) - \frac{\mathbf{a}^T \mathbf{S}^{-2}\mathbf{a}}{2},
\end{align}
which is a quadratic function of the unknown state $\mathbf{a}$.

The reconstruction is obtained by setting the $\mathbf{a}$-derivative to zero:
\begin{align}
	\pdv{\ln p(\mathbf{a}|\mathbf{y})}{\mathbf{a}}=\frac{\mathbf{\mathbf{\Theta}}^T \mathbf{y}}{\eta^2}-\frac{\mathbf{\mathbf{\Theta}}^T \mathbf{\mathbf{\Theta}} \mathbf{a}}{\eta^2}-\mathbf{S}^{-2}\mathbf{a}=0,
\end{align}
which results in a simple linear equation for state reconstruction. Solving the equation, we get the following prescription for reconstruction:
\begin{align}
	\hat{\mathbf{a}}=\left( \mathbf{S}^{-2}+\frac{\mathbf{\mathbf{\Theta}}^T \mathbf{\mathbf{\Theta}}}{\eta^2} \right)^{-1} \frac{\mathbf{\Theta}^T \mathbf{y}}{\eta^2}=\mathbf{A}^{-1}\frac{\mathbf{\Theta}^T \mathbf{y}}{\eta^2},
	\label{eqn:reconstruction}
\end{align}
which combines the information from the prior and the sensors. This reconstruction is linear and works for any values of sensor measurements $\mathbf{y}$, and thus does not say which set of sensors is better or worse and thus does not guide our sensor selection.

The reconstruction depends on inverting the composite matrix $A$, which might be ill-conditioned, meaning that small errors or noise in sensor measurements $\mathbf{y}$ can result in large error in the reconstructed state. We thus need to connect the reconstruction uncertainty to the metrics of matrix condition. Once such a metric is formulated, sensor placement can be designed to optimize it.

\subsection{Reconstruction uncertainty heatmap}
The reconstruction formula \eqref{eqn:reconstruction} gives the maximal likelihood state, but the uncertainty around that state is non-uniformly distributed. In order to quantify the uncertainty, here we compute the uncertainty heatmap across the whole domain of $\mathbf{x}$, given the sensor placement \chgtwo{$\mathbf{C}$}. We denote the sensor reading fluctuation as $\mathbf{\Delta y}$, and propagate that fluctuation to the state reconstruction:
\begin{align}
	\mathbf{\Delta x}=\mathbf{\Psi}_r A^{-1}\frac{\mathbf{\Theta}^T}{\eta^2} \mathbf{\Delta y}.
\end{align}

The state fluctuation depends on the realization of sensor noise, which needs to be averaged out. We can compute the average covariance matrix between all the entries of $\mathbf{\Delta x}$ by taking the outer product of the state fluctuation with itself:
\begin{align}
	\expval{\mathbf{\Delta x} \mathbf{\Delta x}^T}= \mathbf{\Psi}_r \mathbf{A}^{-1} \frac{\mathbf{\Theta}^T}{\eta^2} \expval{\mathbf{\Delta y} \mathbf{\Delta y}^T} \frac{\mathbf{\Theta}}{\eta^2} \mathbf{A}^{-1} \mathbf{\Psi}_r^T,
\end{align}
where the sensor reading covariance is $\expval{\mathbf{\Delta y} \mathbf{\Delta y}^T}=\mathbf{I}_p \eta^2$ by the assumption of uncorrelated noise. The whole state covariance matrix is $n\times n$, which characterizes uncertainty correlations between different locations but does not easily fit in computer memory for large state spaces.

We instead compute only the diagonal part of the covariance matrix, characterizing the level of uncertainty in each pixel of the reconstructed state:
\begin{align}
	\mathbf{B}\equiv& \mathbf{\Psi}_r \mathbf{A}^{-1} \frac{\mathbf{\Theta}^T}{\eta^2}\\
	\mathbf{\sigma}_i =& \eta \sqrt{\sum\limits_j (\mathbf{B}_{ij})^2},
    \label{eqn:uncertheatmap}
\end{align}
where the matrix $\mathbf{B}$ has dimensions $n\times p$ and the resulting vector $\mathbf{\sigma}$ contains the standard deviation of noise in each pixel of the reconstructed image.

\subsection{Sensor energy landscape}
In order to enable systematic design of the sensor configuration, we aim to maximize the determinant of the matrix $\mathbf{A}$ in the reconstruction \eqref{eqn:reconstruction}. The matrix determinant corresponds to the volume of the confidence ellipsoid around the maximal likelihood reconstruction\chg{, as well as the expected Shannon information gain \cite{chaloner1995bayesian}}. The choice to maximize the determinant is known as D-optimal design \cite{deaguiar1995doptimal, manohar2018data}, contrasted with A-optimal and E-optimal designs (\chg{inverse} matrix trace and spectral gap, respectfully).

The general idea of the computation is to relate the determinant of $\mathbf{A}$ to the locations of the sensors, both in absolute space and with respect to each other. The dependence of sensor placement on absolute coordinates is equivalent to a 1-body interaction, or external field. The dependence of sensor placement on relative positions is equivalent to 2-body, 3-body, and higher order sensor interactions.

\chgtwo{In statistical physics, models with these properties are commonly referred to as the \emph{Ising model} originally developed to explain spontaneous magnetization via alignment of magnetic moments of atoms in metals \cite{brush1967history}. The two defining features of an Ising model is a binary nature of each decision variable and the interaction of no more than two variables in a single term, although many variations exist in modern literature \cite{lynn2016maximizing, lynn2019surges}. The sensor placement problem qualitatively appears to have those two features: each pixel either has a sensor or not, and a set of sensors attempts to capture the signal variance while avoiding redundancy between the sensors.} Below we derive the functional form of interactions to all orders directly from the training data \chgtwo{in order to make the mapping from sensor placement to the Ising model explicit and quantitative.}

We start with transforming the determinant of $\mathbf{A}$ into the determinant of a related matrix by using Sylvester's determinant theorem:
\begin{align}
	\det \mathbf{A}&=\det(\mathbf{S}^{-2}+\frac{\mathbf{\Theta}^T\mathbf{\Theta}}{\eta^2})\nonumber\\
	&=\det(\mathbf{S}^{-2})\det(\mathbf{I}+ \mathbf{S}^{2}\frac{\mathbf{\Theta}^T\mathbf{\Theta}}{\eta^2})\nonumber\\
	&=\det(\mathbf{S}^{-2})\det(\mathbf{I}+\frac{\mathbf{\Theta} \mathbf{S}^2 \mathbf{\Theta}^T}{\eta^2}),
	\label{eqn:det}
\end{align}
which converts an $r\times r$ matrix into a $p\times p$ matrix, with size directly related to the number of sensors.

To deepen the analogy with energy in physics, we identify the \emph{negative} log-determinant with the Hamiltonian of a sensor set $\gamma$:
\begin{align}
	\mathcal{H}(\gamma)\equiv-\ln\det(\mathbf{A})=E_b-\Tr\ln(\mathbf{I}+\frac{\mathbf{\Theta} \mathbf{S}^2 \mathbf{\Theta}^T}{\eta^2}),
	\label{eqn:Hgamma}
\end{align}
where we used the identity $\ln\det \mathbf{M} = \tr\ln \mathbf{M}$ for any generic matrix $\mathbf{M}$.

We identify the expression within the logarithm with an outer product of a matrix with itself $\mathbf{\Theta} \mathbf{S}^2 \mathbf{\Theta}^T\equiv \mathbf{G}_\gamma \mathbf{G}_\gamma^T$. We term the row vectors $\mathbf{g}_i,i\in\gamma$ \emph{sensing vectors}; the matrix $\mathbf{G}_\gamma$ is then assembled from a subset of rows of $\mathbf{G}\equiv \mathbf{\Psi}_r \mathbf{S}$ that correspond to chosen sensors $i\in\gamma$. The goal of the subsequent derivation is to relate $\mathcal{H}(\gamma)$ to the selected sensing vectors \chg{(see Supplementary Materials (SM) for derivation)}.

The resulting Hamiltonian \chg{has an identical functional form} to the Ising model found across statistical physics:
\begin{align}
	\mathcal{H}_{2pt}(\gamma)\equiv&-\ln(\det \mathbf{A})\approx E_b+\sum\limits_{i\in \gamma} h_i + \sum\limits_{i\neq j \in \gamma} J_{ij}\label{eqn:Hgamma_resum}\\
	h_i\equiv& -\ln(1+\mathbf{g}_i\cdot \mathbf{g}_i/\eta^2)\leq 0\label{eqn:hi_resum}\\
	J_{ij}\equiv& \frac{1}{2} \frac{(\mathbf{g}_i\cdot \mathbf{g}_j/\eta^2)^2}{(1+\mathbf{g}_i\cdot\mathbf{g}_i/\eta^2)(1+\mathbf{g}_j\cdot\mathbf{g}_j/\eta^2)} \geq 0,\label{eqn:Jij_resum}
\end{align}
where $\mathbf{g}_i$ are the \emph{sensing vectors} describing the sensitivity of each possible sensor location to each of the POD modes, computed as rows of the data-driven matrix $\mathbf{G}=\mathbf{\Psi}_r \mathbf{S}$. The functional form of $h_i$ and $J_{ij}$ is computed via series expansion of the matrix $\mathbf{A}$ in powers of $\eta$ (Eqn.~\ref{eqn:reconstruction}) and resummation, similar to enumeration arguments in self-assembly studies \cite{murugan2015undesired, klishin2021topological}.

What are the limits of \chg{approximating the Hamiltonian with the first two terms}? The answer to this question is intimately tied to the sensor placement algorithm. Generically, we expect the approximation to work while $\mathbf{g}_i\cdot \mathbf{g}_j\ll \mathbf{g}_i\cdot\mathbf{g}_i$, i.e. the correlation between the sensing vectors is small compared with their magnitude. \chg{We thus formulate two predictions of approximating Eqn.~\ref{eqn:Hgamma} with Eqn.~\ref{eqn:Hgamma_resum}: (i) the approximation should work better for small number of sensors and (ii) the approximation should work better for near-optimal sensor configurations rather than generic ones. The first prediction is driven by the \emph{number} of crosstalk terms that grows as $p^2$ for $p$ sensors, and the second is driven by the \emph{magnitude} of the crosstalk terms since the sensor placement methods aim to minimize it. The predictions can only be tested once training data is available and the landscapes $h_i,J_{ij}$ take specific values, thus we test them in Section IV.B and Fig.~\ref{fig:methods} below.}


We note that the Hamiltonian terms that involve more than two sensors (three, four, etc.) would have the shape similar to $J_{ij}$ in Eqn.~\ref{eqn:Jij_resum}, with a large number of indices. Due to the construction of the series expansion via a non-diagonal matrix, the terms where \emph{adjacent} indices are identical would vanish. However, the indices can repeat in non-adjacent positions, e.g. at fourth order in $J^{(4)}_{ijij}\neq 0$. The 2-point expression Eqn.~\ref{eqn:Hgamma_resum} is thus not exact even for placement of 2 sensors, but is expected to be a good approximation.

\subsection{High and low noise limits}
Here we consider the high noise and low noise limits of the landscape \eqref{eqn:hi_resum},\eqref{eqn:Jij_resum}. In the high noise limit $\eta\gg 1$ we get:
\begin{align}
	h_i=&\order{\eta^{-2}}\\
	J_{ij}=&\order{\eta^{-4}},
\end{align}
so the crosstalk falls off faster than the 1-sensor landscape. On one side, this stimulates putting more sensors in the basin of lowest $h_i$: since sensor noise is high, it makes more sense to collect measurements in the location of highest signal variance. On the other side, placing sensors close by breaks the approximation condition $\mathbf{g}_i\cdot \mathbf{g}_j\ll \mathbf{g}_i\cdot\mathbf{g}_i$, leading to a faster divergence between the 2-point energy and the true energy.

In the low noise limit $\eta\ll 1$:
\begin{align}
	h_i=&2\ln(\eta)-\ln(\mathbf{g}_i \cdot\mathbf{g}_i)\\
	J_{ij}=&\frac{1}{2} \frac{(\mathbf{g}_i\cdot \mathbf{g}_j)^2}{(\mathbf{g}_i\cdot\mathbf{g}_i)(\mathbf{g}_j\cdot\mathbf{g}_j)},
\end{align}
which approaches a constant, noise-independent shape where neither the 1-point nor 2-point or higher order interactions vanish for a generic sensor set. It should still be possible to choose a small set of sensors with low crosstalk and ensure that the 2-point energy is a good approximation of the true energy. Importantly, in low noise limit the sensor placement landscape does not depend on the absolute magnitude of the prior, but does depend on the its shape, i.e. uniform and non-uniform priors would typically result in different landscapes and thus different chosen sensor sets.

\subsection{Sensor placement methods}
\label{sec:methods}

In the Hamiltonian formulation of sensor placement Eqn.~\ref{eqn:Hgamma_resum}, the objective depends on the locations of individual sensors and sensor pairs from the chosen set $\gamma$ (Fig.~\ref{fig:landscape}). Qualitatively, minimizing the Hamiltonian requires picking sensors $i\in\gamma$ that capture a lot of signal variance (large $\mathbf{g}_i\cdot\mathbf{g}_i$), but are not very correlated with each other (small $\mathbf{g}_i\cdot\mathbf{g}_j$). While a combinatorial search for the lowest energy configuration would require evaluating the $\order{n^2}$ elements of the full crosstalk $\mathbf{J}$ matrix, we consider and compare several greedy placement methods of sensor placement:
\begin{enumerate}
	\item \textit{Random method}: sensors are placed uniformly randomly within the signal domain without overlap.
	\item \textit{1-point method}: sensors are placed to greedily minimize the 1-point energy $h_q$ without overlap:
	\begin{align}
		q=\arg\min\limits_{q\notin \gamma} h_q;\; \gamma\gets q, \label{eqn:method1pt}
	\end{align}
	requiring just a single evaluation of the energy landscape at cost \chgtwo{$\order{nr}$ multiplications} to place any number of sensors $p$. 
	\item \textit{2-point method}: sensors are placed to greedily minimize the 2-point energy without overlap:
	\begin{align}
		q=\arg\min\limits_{q\notin \gamma} \left(h_q+2\sum\limits_{i\in \gamma} J_{iq}\right);\; \gamma\gets q, \label{eqn:method2pt},
	\end{align}
	requiring just \chgtwo{$\order{nrp}$} evaluations to place $p\ll n$ sensors.
	\item \textit{QR method}: \chgtwo{$p=r$} sensors are placed via greedy QR factorization of the $\mathbf{\Psi}_r$ matrix with the \texttt{PySensors} package \cite{desilva2021pysensors} \chgtwo{at the cost of $nr^2$ multiplications}.
\end{enumerate}

\chgtwo{All computational complexity calculations focus on the number of arithmetic multiplications required, as opposed to arithmetic additions or evaluations of $\arg \min$ of a \emph{known} fixed-size array.}

%

\section{Results}
\label{sec:results}
\subsection{Sensor placement landscapes}
\begin{figure}
	\begin{center}
		\includegraphics[width=\columnwidth]{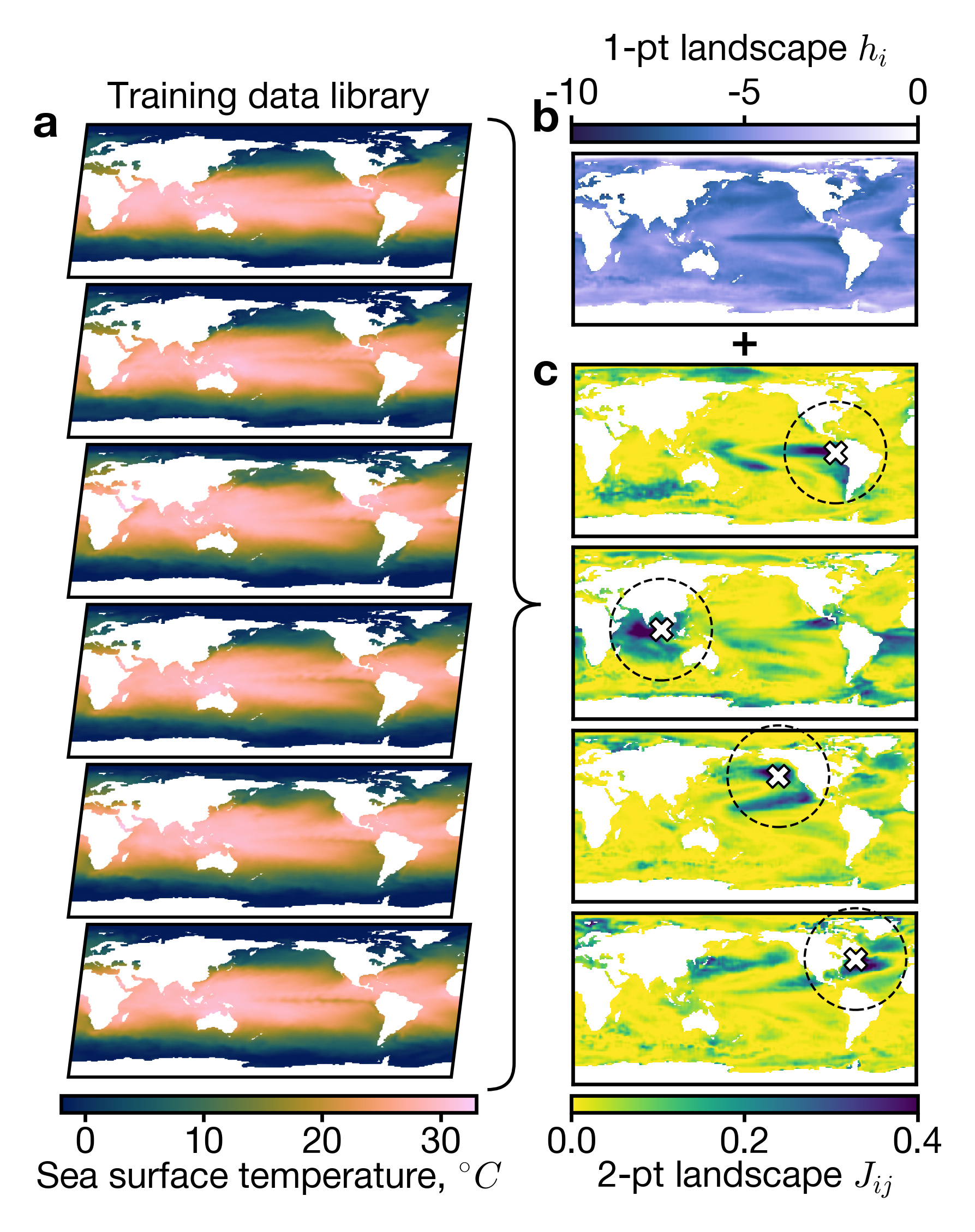}
	\end{center}
	\caption{Computation of the sensor placement landscape from the training data library. (a) A collection of data snapshots of identical dimensions, here the Sea Surface Temperature data set. (b) The library is used to compute the sensor placement landscape to multiple orders, here showing the 1-point landscape at each location $h_i$ and the 2-point landscape $J_{ij}$ conditioned on the sensor locations marked with white crosses and highlighted with dashed circles.}
	\label{fig:landscape}
\end{figure}

We now use the derived energy landscape expressions \eqref{eqn:hi_resum}-\eqref{eqn:Jij_resum} to illustrate the shape of the landscapes for a specific dataset of weekly average sea surface temperature (SST) between 1990 and 2023 \cite{huang2021improvements}, truncated to POD rank $r=100$. Each frame covers the entirety of Earth surface in equirectangular projection at $1^\circ$ resolution, resulting in $360\times 180$ pixel images with $n=44219$ pixels corresponding to sea surface (Fig.~\ref{fig:landscape}a). The preprocessing and centering of the training library  is described in the SM.

The 1-point landscape $h_i$ is computed directly from the training library and shown in the same coordinates in Fig.~\ref{fig:landscape}b. The 1-point landscape captures the amount of signal variance at each location, with the locations of highest variance being preferred and shown in the darkest shade in the figure. While the landscape computation algorithm has no input of ocean and atmospheric physics and acts on a collection of vectorized state snapshots, the resulting landscape readily identifies the major geographical features of Earth. In particular, $h_i$ is low in inland bodies of water such as the Baltic, Black, and Caspian seas in Eurasia, and the Great Lakes and Hudson's Bay in North America, as well as close to the continental coasts~\chgtwo{ and in the El Niño region of the Pacific Ocean along the equator.}

\begin{figure*}[ht!]
	\begin{center}
		\includegraphics[width=0.9\textwidth]{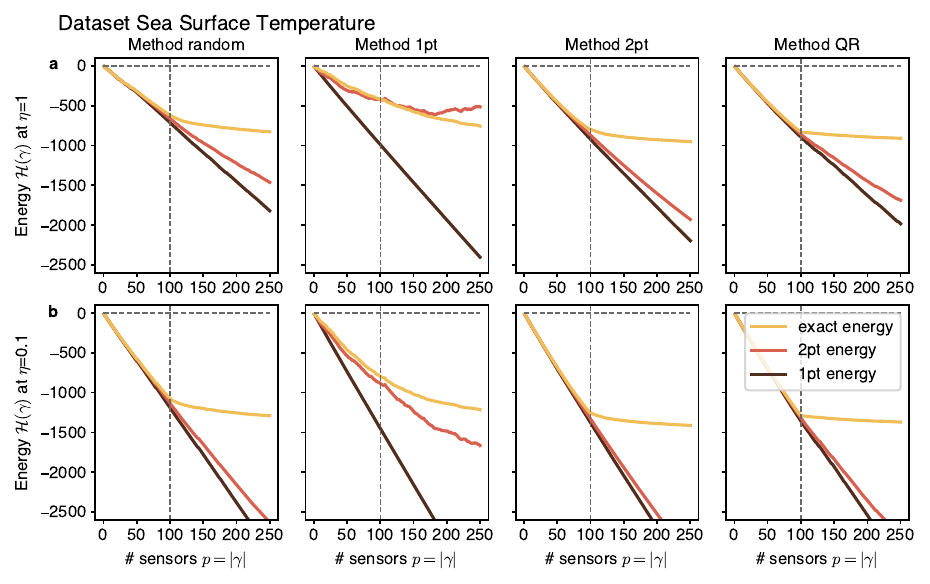}
	\end{center}
	\caption{Comparison of sensor placement methods for the \chg{Sea Surface Temperature} dataset as measured by different energy expressions. The rows (a) and (b) use the energy landscape at different values of noise $\eta$ \chgtwo{($0.5 \sigma_{scale}$ and $0.05\sigma_{scale}$, respectively)}, the columns correspond to four different sensor placement methods. The curves correspond to three energy formulae. Exact energy is evaluated by directly computing the determinant in Eqn.~\ref{eqn:Hgamma}, ignoring the constant term $E_b$. 2-point energy is evaluated with Eqn.~\ref{eqn:Hgamma_resum}. 1-point energy is evaluated by taking only the first term in Eqn.~\ref{eqn:Hgamma_resum}. The horizontal dashed line indicates $\mathcal{H}=0$, the vertical line indicates the reconstruction rank $p=r$.}
	\label{fig:methods}
\end{figure*}

The 2-point landscape $J_{ij}$ is also computed directly from the training library. Since the sensor interaction (cross-talk) can be computed for any pair of sensor locations, in full $J_{ij}$ is an $n\times n$ matrix that is hard to fit in memory or visualize. By the formula \eqref{eqn:Jij_resum}, $J_{ij}\geq 0$, so the 2-point sensor interaction can never decrease the energy of any sensor configurations. In other words, sensor interactions can make sensing worse, but never better. In order to show the magnitude of this effect quantitatively, we visualize the 2-point landscape in a conditional form in Fig.~\ref{fig:landscape}c. As an example, placing a sensor in the Pacific Ocean off the coast of South America penalizes the placement of more sensors in the nearby domain and some remote areas. Similarly, placing a sensor in North Indian Ocean penalizes other sensors nearby. While the pattern of sensor interactions is complex, it is mostly local, with far away sensors having no effect on each other, reflecting the mostly local nature of ocean and atmospheric dynamics. \chgtwo{Sensor placement landscapes for two other empirical datasets are presented in SM and demonstrate different domain-specific features.}

\subsection{Sensor placement methods comparison}

While the visualization of the energy landscapes helps the qualitative interpretation of sensor placement, it remains important to check whether the 2-point landscape computation enables a good sensor placement. The exact energy expression \eqref{eqn:Hgamma} offers a way to evaluate any sensor set, but not to select a set. We thus compare the four methods of sensor placement listed in Section \ref{sec:methods} by placing sensors in sequence and computing both the exact energy and the 1-point and 2-point approximations.

We consider sensor placement methods with \chgtwo{the same SST dataset.}
The goal of all four methods is to minimize the sensor configuration energy $\mathcal{H}(\gamma)$ by minimizing different proxies (Fig.~\ref{fig:methods}). Across all methods, exact energy is higher than 1-point energy, indicating the importance of taking sensor crosstalk into account. For the 2-point and QR methods, the 2-point energy is a close approximation for the exact energy up until the number of sensors reaches the reconstruction rank $p=r$. For the 1-point method, the discrepancy between the exact and 2-point energies is the highest since placing the sensors without considering crosstalk results in strong spatial clustering and thus large crosstalk. The random placement method has performance better than 1-point and worse than 2-point and QR.


\subsection{Reconstruction progress}
\begin{figure}
	\begin{center}
		\includegraphics[width=\columnwidth]{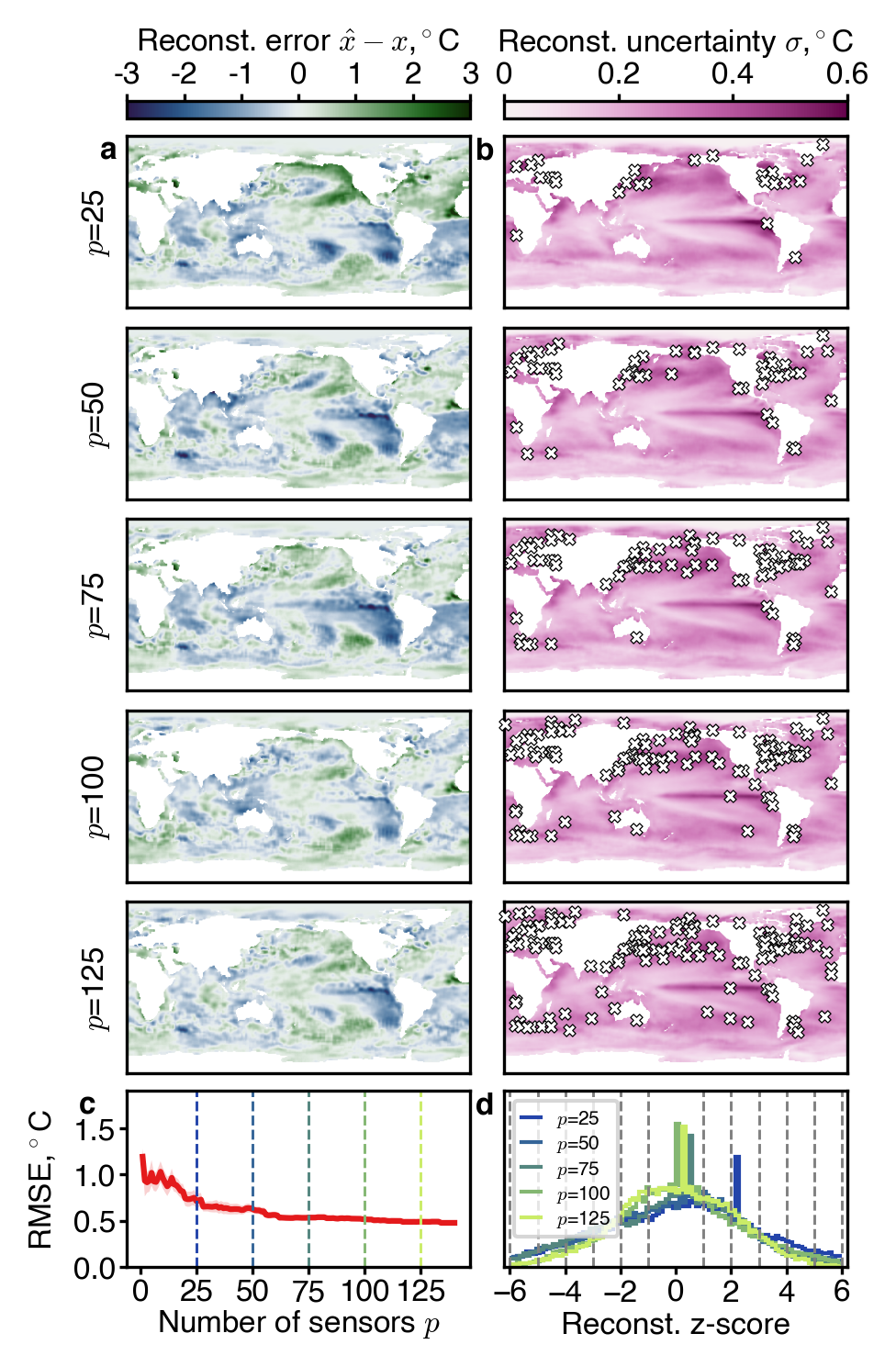}
	\end{center}
	\caption{State reconstruction with progressively more sensors $p$ at noise level $\eta=1.0^\circ$ C. (a) Reconstruction error between the maximal likelihood reconstructed state and the true state. (b) Uncertainty heatmaps of the reconstruction and the sensor locations chosen by the 2-point algorithm (white crosses). (c) RMSE of the reconstruction with the solid curve and the shaded region showing mean $\pm$ one standard deviation across 20 realizations of sensor noise. The RMSE peak corresponds to the model dimension, here $r=100$. (d) Distribution of reconstruction z-scores at every pixel $z_i=(\hat{x}_i-x_i)/\sigma_i$. \chgtwo{The standard deviations of the z-scores are [3.61, 3.30, 2.16, 2.37, 2.01] for 25...125 sensors, respectively}}
	\label{fig:reconstruction}
\end{figure}

\color{black}
Having established that the 2-point method is closely comparable to the QR method, we now turn to analysis of state reconstruction on the SST dataset. We show that by employing as few as 25 sensors selected by the 2-point algorithm with noise level of $\eta=1^\circ$ C, the entire temperature field can be reconstructed to within $1^\circ$ C (Fig.~\ref{fig:reconstruction}a). The reconstruction method also provides an Uncertainty Quantification (UQ) method in form of the uncertainty heat map at every pixel (\chgtwo{computed in Eqn.~\ref{eqn:uncertheatmap} and visualized in }Fig.~\ref{fig:reconstruction}b). \chgtwo{Uncertainty is high in the same regions where signal variance is high, including inland bodies of water, continental coasts, and the El Niño region along the equator in the Pacific Ocean, and low in mid-ocean regions far from coasts.}
We emphasize that the sensor placement algorithm was trained exclusively on the snapshot library, with no additional information about the structure of Earth's oceans or their physical processes.

\chgtwo{The reconstruction Root Mean Square Error (RMSE) shows monotonic decrease with the number of sensors (Fig.~\ref{fig:reconstruction}c) along the whole axis. While adding more sensors contributes more information to the reconstruction algorithm, it also trades off with the number of independent sources of noise, resulting in the curve flattening for $p>50$. In order to compare the reconstruction error and the prediction uncertainty, we assess the model confidence by computing the z-score of each individual pixel $z_i=(\hat{x}_i-x_i)/\sigma_i$ and plotting its distribution (Fig.~\ref{fig:reconstruction}d). Whereas for a properly calibrated UQ, the z-score is supposed to have a standard deviation close to 1, in our case the distributions are wider with standard deviations ranging from 3.61 to 2.01. We conclude that while the Eqn.~\ref{eqn:uncertheatmap} correctly captures the \emph{order of magnitude} of uncertainty for the first time, it misses important error contributions from other sources.}


\subsection{Reconstruction diagnostics}
\begin{figure*}
	\begin{center}
		\includegraphics[width=0.95\textwidth]{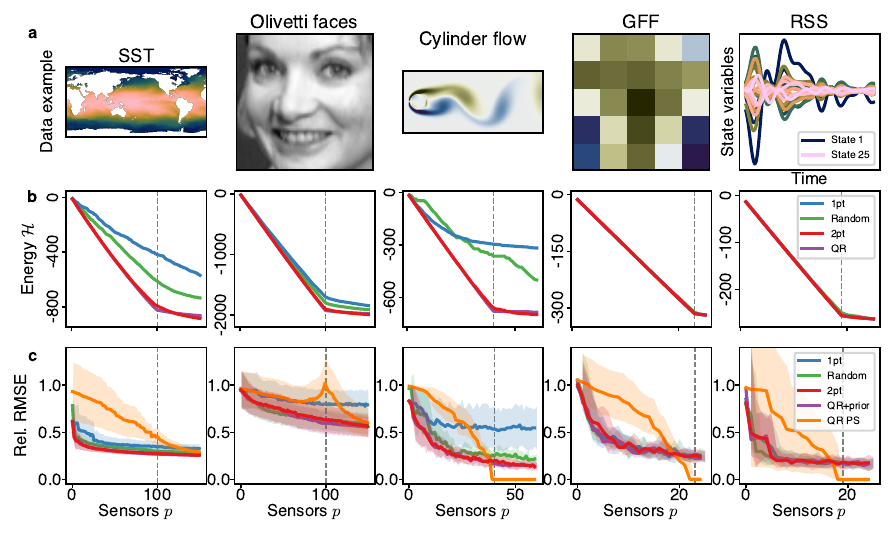}
	\end{center}
	\caption{Comparison of reconstruction metrics for different datasets, with 4:1 \chgtwo{randomized} train/test split. (a) Examples of dataset snapshots for different systems. The RSS dataset is non-spatial, so temporal trajectories are shown instead. (b) Exact energies of sensor configurations selected with different methods: random placement, 1-point algorithms, 2-point algorithm, and QR pivoting. (c) \chgtwo{Relative root mean square error} (RMSE) of state reconstruction \chgtwo{for sensor readings corrupted with noise of $\eta=0.5\sigma_{scale}$}, with solid line and shaded region indicating average $\pm$ 1 standard deviation across the test set. The vertical dashed line indicates the number of modes $r$ used in the reconstruction.}
	\label{fig:comparison}
\end{figure*}

We compute sensor placement and reconstruction error across five datasets and four sensor placement methods. Apart from the SST dataset, we use the Olivetti faces dataset \cite{samaria1994parameterisation}, snapshots of a numerical simulation of flow past a cylinder \cite{taira2007immersed, colonius2008fast}, as well as synthetic Gaussian Free Field \cite{Cadiou2022Fyeld} and Random State System \cite{fuller2021python} datasets (Fig.~\ref{fig:comparison}a, see SM for dataset details). \chgtwo{In order to facilitate the comparison between the datasets, we compute the data scale $\sigma_{scale}$, defined as the standard deviation across all elements of the centered training data matrix (listed for each dataset in the SM). We use the data scale twice: to set the magnitude of sensor noise $\eta$, and to normalize the reconstruction error.}
For the QR sensors, we compute reconstruction with and without the prior regularization. For each dataset, \chgtwo{the POD modes and thus the sensor placement landscape are derived from the randomized} training set with 80\% of the data, and the reconstruction error is computed across the test set with the remaining 20\% of the data.

For the three empirical datasets (SST, Olivetti, and cylinder) the random and 1-point algorithms have higher energies than the other two (Fig.~\ref{fig:comparison}b). The 1-point algorithm has higher RMSE than other reconstructions with prior, highlighting the importance of crosstalk for sensor placement (Fig.~\ref{fig:comparison}c). The QR sensors without prior regularization also show consistently higher RMSE, justifying the need for a prior. The regularized random, 2-point, and QR algorithms show nearly equivalent RMSE error curves \chgtwo{in contrast with the legacy unregularized QR PS algorithm that shows the peak at $p\sim r$ due to the POD mode truncation}
For the two synthetic datasets all sensor placement methods have nearly equivalent performance \chgtwo{with regularized reconstruction; for these small datasets, algorithmic placement} can also be compared to brute force search (see SM). We conclude that while the 2-point and the QR algorithms are based on the same underlying POD modes and have nearly equivalent numerical performance, the 2-point algorithm provides much richer interpretation in terms of sensor landscapes and interactions.

\section{Discussion and outlook} 
\label{sec:discussion}
The \chgtwo{first} key advance of this paper is casting the sensor selection problem in statistical mechanics terms of interaction energies of progressively larger numbers of sensors. While we focus the discussion on the 1-body and 2-body interactions, the mathematical formalism extends to any higher number (see SM). The shape of the 2-body interactions can be further connected to the properties of the physical, mathematical, or even artistic processes that generate data \cite{stephens2013statistical, duplantier2017log, kentdobias2022log}. We used a greedy 2-point method of sensor placement in order to limit the required memory and computing time, but if the whole landscape could fit in memory, better energy minima can be obtained through methods such as gradient descent or simulated annealing \cite{kirkpatrick1983optimization}. Due to the usage of a regularizing prior, state reconstruction can be consistently performed for any number of sensors without the requirement that $p\geq r$ \cite{manohar2018data}.

\chgtwo{The qualitative and quantitative connection between sensor placement algorithms to the Ising model has two main advantages. On the one hand, sensor placement landscapes can be directly visualized and interpreted in the data domain as shown in Fig.~\ref{fig:landscape} for SST and in Figs.~S5-S6 for Olivetti faces and cylinder flow. These landscapes highlight where the sensors ``want'' to be placed and how they ``repel'' each other in an \emph{a priori} fashion before any sensors were placed and independently of a sensor placement algorithm. On the other hand, computing the energy of \emph{any} (not necessarily optimal) sensor set up to 2-point functions can be done in $\order{p^2}$ operations as opposed to $\order{p^3}$ for the determinant-based exact energy. Updating individual sensor positions within the same set can be done even cheaper, opening the way for \emph{post hoc} refinement of a sensor placement determined by one of the greedy algorithms presented here.}

\chgtwo{The second key advance of this paper is the regularized reconstruction of Eqn.~\ref{eqn:reconstruction} that significantly decreases the reconstruction RMSE (risk curve) in the undersampled regime $p<r$.}
\chg{For some datasets, the \chgtwo{risk curve} shows a seemingly counter-intuitive feature of instability at $p\approx r$ (\chgtwo{e.g. the Olivetti faces dataset in }Fig.~\ref{fig:comparison}c):} that is, adding \emph{more} sensors can make the quality of reconstruction significantly \emph{worse} before it gets better again. \chgtwo{Such behavior has been observed in sparse sensing plots before but never commented upon \cite{clark2018greedy, desilva2021pysensors}.} \chgtwo{Elsewhere in the literature, this} phenomenon is known as \emph{double descent} and has recently attracted significant attention in deep learning studies \cite{belkin2019reconciling}, but has been noticed in linear classifiers even back in the 1980s \cite{loog2020brief}. In supervised learning studies, double descent has been observed when the number of model parameters matches the number of data points, and thus the model becomes extremely sensitive to each data point. In our case of sparse sensing, if double descent is observed at all, the risk curve peaks when the number of sensors matches the number of modes.

The general shape of the double descent curve can be quantitatively predicted with tools from statistical physics of disordered systems \cite{dascoli2020double}, and the peak can be mitigated with an appropriate choice of reconstruction regularization \cite{nakkiran2020optimal}.
However, those arguments rely on averaging over the random features, whereas here we aim to choose specific sensor locations deliberately.  \chgtwo{Our results in this paper show that regularized reconstruction suppresses the double descent phenomenon in sensor placement context, but do not exhaustively explain the shape of the risk curve across the datasets.} The connection from sensor placement landscapes and algorithms to the shape of risk curve\chgtwo{, as well as the optimal design of reconstruction regularization remain important questions} for further work.

While the sensor landscape approach provides interpretability and in some regimes selects better sensor sets than state-of-the-art approaches, it is ultimately followed by a \emph{linear} algorithm for reconstructing the state from sensor readings, limiting the reconstruction accuracy. Recent work has shown that a linear algorithm of sensor selection based on QR factorization can be combined with a nonlinear shallow decoder network state estimation, which nevertheless requires neural network retraining for any new sensor set \cite{williams2022data}. An alternative approach instead focuses on learning the data manifold geometry and identifying nonlinear coordinates \cite{otto2022inadequacy}, which would be equivalent to replacing the Gaussian prior in our approach with a more complex one. Other sensor placement extensions can involve estimation of time-dependent dynamics through Kalman filtering \cite{tzoumas2016sensor}, or sensors advected by the flows they are trying to measure \cite{shriwastav2022dynamic}. Finally, the approach here identifies only the part of sensor placement landscape induced by the training data, which can be combined with other design objectives such as placement cost or restrictions \cite{klishin2018statistical, clark2018greedy, nishida2022region}. \chgtwo{The uncertainty heatmap formula of Eqn.~\ref{eqn:uncertheatmap} has been used in nuclear reactor applications \cite{karnik2024optimal, karnik2024energies}.}

\chgtwo{The computational methods reported in this paper have been incorporated into the \texttt{PySensors} package \url{https://github.com/dynamicslab/pysensors}.}

\section*{Acknowledgments}
The authors would like to thank S.E.~Otto and J.~Williams for helpful discussions and L.D.~Lederer for administrative support. This work uses Scientific Color Maps for visualization \cite{crameri2023color}. The authors acknowledge support from the National Science Foundation AI Institute in Dynamic Systems (grant number 2112085).

\bibliography{sensors}

\end{document}


\title{Data-Induced Interactions of Sparse Sensors Using Statistical Physics: \\ Supplementary Materials}
	\author{Andrei A.~Klishin}
        \email{aklishin@hawaii.edu}
	\author{J.~Nathan Kutz}
	\author{Krithika Manohar}
        \email{kmanohar@uw.edu}
	
	\date{\today}
	
	\maketitle

\section{Data preprocessing}
\subsection{Sea Surface Temperature}
The Sea Surface Temperature (SST) dataset is curated by the National Oceanic and Atmospheric Administration (NOAA) and employs the Optimal Interpolation (OI) method to collect observations from different platforms and places them on a single regular grid for long periods of observations \cite{huang2021improvements}. In this work we use OISST 2.0 dataset \footnote{At the time of writing the dataset was available via the Physical Sciences Laboratory at \url{https://psl.noaa.gov/data/gridded/data.noaa.oisst.v2.html}.} that has the spatial resolution of 1 degree in both latitude and longitude, and is temporally averaged for each week between Dec 31, 1989 and Jan 1, 2023 for a total of $N=1727$ snapshots. Each temperature snapshot is given in equirectangular projection of size $360\times 180$ pixels. All of land surface of Earth is excluded from observations via a time-independent binary mask, leaving $n=44219$ pixels that vary between the states. We center the dataset by subtracting the temporal mean temperature profile from each snapshot. The POD is truncated to rank $r=100$. \chgtwo{Natural scale of the dataset is $\sigma_{scale}=1.945^\circ C$.}

\subsection{Olivetti faces}
The Olivetti faces dataset consists of photos of individuals taken between April 1992 and April 1994 at AT\&T Laboratories Cambridge \cite{samaria1994parameterisation}. Each of 40 individuals has 10 images, for a total of $N=400$ presented in random order. Each photo has the size $64\times 64$ grayscale pixels, resulting in $n=4096$. We center each image both locally and globally by subtracting the mean \chgtwo{face brightness profile from each image}. The POD is truncated to rank $r=100$. \chgtwo{Natural scale of the dataset is $\sigma_{scale}=0.1231$ (dimensionless).}

\subsection{Cylinder flow}
The cylinder flow dataset consists of scalar vorticity fields in 2D flow of a liquid around a cylinder, numerically simulated with immersed boundary method in Refs.~\cite{taira2007immersed, colonius2008fast}. In the simulation regime at Reynolds number $Re=100$, the flow consists of periodic shedding of vortices from the two sides of the cylinder in alternating order. The data consists of $N=151$ snapshots of resolution $449\times 199$ pixels, resulting in data dimension $n=89351$. We center the data by subtracting the temporal average vorticity profile from each snapshot. The POD is truncated to rank $r=40$. \chgtwo{Natural scale of the dataset is $\sigma_{scale}=0.550$ (dimensionless).}

\subsection{Gaussian Free Field}
The Gaussian Free Field (GFF) dataset is generated synthetically with the FyeldGenerator package \cite{Cadiou2022Fyeld}. The field is generated by drawing pseudorandom values of Fourier modes according to a user-specified power spectrum. In order to ensure that the field has coarse features on small spatial scale, we chose power spectrum $P(k)\propto k^{-10}$ as opposed to the usual $P(k)\propto k^{-2}$ for standard GFF. We generated $N=50$ pseudorandom snapshots of a field in the spatial domain $5\times 5$ pixels, resulting in $n=25$. The small size of the synthetic dataset was required to enable the exhaustive enumeration of sensors, see Sec.~\ref{sec:exhaustive}. We center each image both locally and globally by subtracting the mean brightness value of each image and the mean value across the dataset. The POD is truncated to rank $r=23$, equal to the number of singular values above machine precision $\sigma>10^{-15}$. \chgtwo{Natural scale of the dataset is $\sigma_{scale}=247.434$ (dimensionless).}

\subsection{Random State System}
The Random State System (RSS) dataset is generated synthetically with the Python Control Package \cite{fuller2021python} that replicates the functionality of Matlab Control Toolbox. RSS is a linear dynamical system with equation $\dot{\mathbf{x}}=M\mathbf{x}$ for a random square matrix $M$ of specified dimension, which we choose to be $n=25$. The generating package does not request any other free parameters such as the distribution of the random matrix, other than ensuring the dynamical system stability, i.e. $\forall\lambda: Re(\lambda)\leq 0$. We obtain the trajectory data by initializing random iid initial condition $x_i(t=0)\sim \mathcal{N}(0,1)$ and integrating the dynamical equation numerically in the range $t\in[0,5]$, sampling $N=101$ snapshots. The small size of the synthetic dataset was required to enable the exhaustive enumeration of sensors, see Sec.~\ref{sec:exhaustive}. We center each state both locally and globally by subtracting the mean brightness value of each image and the mean value across the dataset. The POD is truncated to rank $r=19$, equal to the number of singular values above machine precision $\sigma>10^{-15}$. \chgtwo{Natural scale of the dataset is $\sigma_{scale}=2.980$ (dimensionless).}

\subsection{Dataset singular values}
\begin{figure*}
	\includegraphics[width=\textwidth]{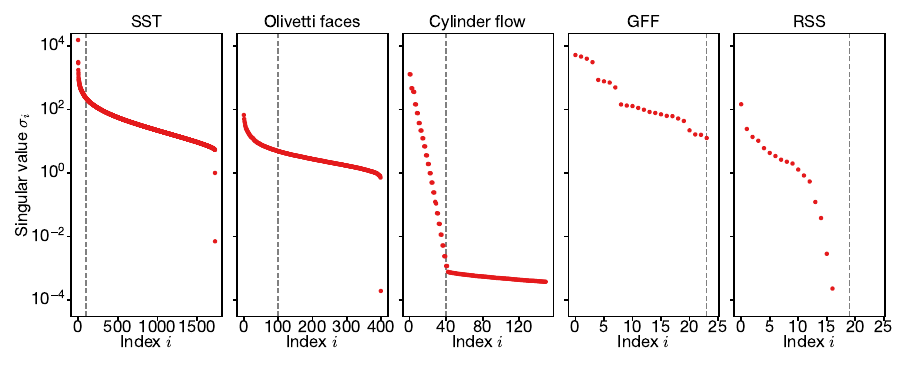}
	\caption{Singular values of the five datasets used. Gray dashed line indicates the POD truncation threshold chosen for each of the datasets.}
	\label{fig:singular}
\end{figure*}
Fig.~\ref{fig:singular} shows the spectra of singular values for each of the five datasets, along with the truncation threshold. As expected, singular values span many orders of magnitude for both empirical and synthetic datasets.


\section{Sensor energy landscape}

\subsection{Determinant decomposition}
To enable systematic design of the sensor configuration, we aim to maximize the determinant of the matrix $A$ in the reconstruction (Eqn.~9 of main text). The matrix determinant corresponds to the volume of the confidence ellipsoid around the maximal likelihood reconstruction. The choice to maximize the determinant is known as D-optimal design \cite{deaguiar1995doptimal, manohar2018data}, contrasted with A-optimal and E-optimal designs (matrix trace and spectral gap, respectfully).

The general idea of the computation is to relate the (log-) determinant of $A$ to the locations of the sensors, both in absolute space and with respect to each other. The dependence of sensor placement on absolute coordinates is equivalent to a 1-body interaction, or external field. The dependence of sensor placement on relative positions is equivalent to 2-body, 3-body, and higher order sensor interactions. Below we derive the functional form of interactions to all orders directly from the training data.

We start with transforming the determinant of $\mathbf{A}$ into the determinant of a related matrix by using Sylvester's determinant theorem:
\begin{align}
	\det \mathbf{A}&=\det(\mathbf{S}^{-2}+\frac{\mathbf{\Theta}^T\mathbf{\Theta}}{\eta^2})\nonumber\\
	&=\det(\mathbf{S}^{-2})\det(\mathbf{I}+ \mathbf{S}^{2}\frac{\mathbf{\Theta}^T\mathbf{\Theta}}{\eta^2})\nonumber\\
	&=\det(\mathbf{S}^{-2})\det(\mathbf{I}+\frac{\mathbf{\Theta} \mathbf{S}^2 \mathbf{\Theta}^T}{\eta^2}),
	\label{eqn:det}
\end{align}
which converts an $r\times r$ matrix into a $p\times p$ matrix, with size directly related to the number of sensors.

To deepen the analogy with energy in physics, we identify the \emph{negative} log-determinant with the Hamiltonian of a sensor set $\gamma$:
\begin{align}
	\mathcal{H}(\gamma)\equiv-\ln\det(\mathbf{A})=E_b-\Tr\ln(\mathbf{I}+\frac{\mathbf{\Theta} \mathbf{S}^2 \mathbf{\Theta}^T}{\eta^2}),
	\label{eqn:Hgamma}
\end{align}
where we used the identity $\ln\det X = \tr\ln X$ for any generic matrix $X$.

We identify the expression within the logarithm with an outer product of a matrix with itself $\mathbf{\Theta} \mathbf{S}^2 \mathbf{\Theta}^T\equiv \mathbf{G}_\gamma \mathbf{G}_\gamma^T$. We term the row vectors $\mathbf{g}_i,i\in\gamma$ \emph{sensing vectors}; the matrix $\mathbf{G}_\gamma$ is then assembled from a subset of rows of $\mathbf{G}\equiv \Psi_r \mathbf{S}$ that correspond to chosen sensors $i\in\gamma$. The goal of the subsequent derivation is to relate $\mathcal{H}(\gamma)$ to the selected sensing vectors.

\subsection{Expansion in $\eta$ and resummation}
The Hamiltonian expression \eqref{eqn:Hgamma} requires taking a matrix logarithm of a complex matrix expression. The outer product $\mathbf{G}_\gamma \mathbf{G}_\gamma^T$ is a positive-semidefinite matrix, and thus the argument of the logarithm $\mathbf{I}+\mathbf{G}_\gamma \mathbf{G}_\gamma^T/\eta^2$ is a positive-definite matrix $\forall \eta$, and the logarithm always exists. Additionally, for a sufficiently large value of $\eta$ the logarithm can be represented as an absolutely and uniformly convergent series expansion in $1/\eta^2$. The strategy of the derivation is thus as follows: (i) assume $\eta$ to be large, (ii) rewrite the energy function as a power expansion in orders of $1/\eta^2$, (iii) perform series resummation into a different closed-form function, (iv) expand the validity of the new function to arbitrary $\eta> 0$ via analytic continuation.

\subsection{Diagonal separation}
We decompose the sensor-driven perturbation as a sum of two matrices:
\begin{align}
	\mathbf{G}_\gamma \mathbf{G}_\gamma^T\equiv \mathbf{D}+\mathbf{R},
	\label{eqn:GGT}
\end{align}
where $\mathbf{D}$ contains only the diagonal elements of the outer product, and $\mathbf{R}$ contains all non-diagonal elements. Since $\mathbf{G}_\gamma \mathbf{G}_\gamma^T$ is a positive-semidefinite matrix, its diagonal $\mathbf{D}$ inherits the same property. Importantly, the two matrices in this decomposition do not commute $[\mathbf{D},\mathbf{R}]\neq 0$, and thus the order of their product is important.

In terms of these newly-defined matrices the Hamiltonian takes the following shape:
\begin{align}
	\mathcal{H}=E_b-\Tr\ln(\mathbf{I}+\frac{1}{\eta^2}(\mathbf{D}+\mathbf{R})),
\end{align}
where $E_b\equiv -\Tr\ln(\mathbf{S}^{-2})$ is the baseline energy independent of noise and sensor choices. We then expand the matrix in powers of $1/\eta^2$:
\begin{align}
	\mathcal{H}(\gamma)=E_b+\Tr\sum\limits_{k=1}^{\infty} \eta^{-2k}(\mathbf{D}+\mathbf{R})^k \frac{(-1)^k}{k},
	\label{eqn:EexactDR}
\end{align}
where we pulled one factor of $(-1)$ outside of the sum. The sum $(\mathbf{D}+\mathbf{R})^k$ cannot be expanded as the simple binomial formula, because the matrices \chg{do not} commute $[\mathbf{D},\mathbf{R}]\neq 0$. Instead, the sum involves many products of $\mathbf{D},\mathbf{R}$ in different order with order-dependent values.

We search for an expression for the Hamiltonian in the following form:
\begin{align}
	\mathcal{H}(\gamma)=E_b+\sum\limits_{s=0}^{\infty}f_s(\mathbf{D}, \mathbf{R}),
\end{align}
where $f_s(\mathbf{R})$ is a function in which $\mathbf{R}$ occurs exactly $s$ times. We can get the form of $f_s$ by grouping terms with the same number of occurrences of $\mathbf{R}$ in the full sum of Eqn.~\ref{eqn:EexactDR}. We treat separately the cases of $s=0$ and $s>0$.

For $s=0$, we want to gather the terms in which the crosstalk matrix $\mathbf{R}$ never occurs. At each order in $k$, there is exactly one such term $\mathbf{D}^k$, which we can resum for all orders of $k$ with appropriate series prefactors:
\begin{align}
	f_0=\Tr\sum\limits_{k=1}^{\infty}\eta^{-2k} \mathbf{D}^k  \frac{(-1)^k}{k}=-\Tr\ln(\mathbf{I}+\frac{1}{\eta^2}\mathbf{D}),
\end{align}
which always exists because $\mathbf{D}$ is positive-semidefinite.

\subsection{Case $s>0$}

Now consider the terms for $s>0$. Omitting the scalar factors, an example would be $\Tr(\mathbf{DRDDRRDDRD})$ where $s=4$. Note that by the cyclic property of the trace, the sequence can be shifted by repeatedly moving terms from the right end to the left end of the product without changing the value of the trace. Several different sequences then contribute the same value to the energy, and the number of such sequences needs to be carefully computed. For the term of order $k$ with $s$ occurrences of the matrix $\mathbf{R}$ there are $\pmqty{k \\ s}$ terms, but not all of them have the same value.

We can write each contributing term in the following form:
\begin{align}
	\Tr(\prod\limits_{i=1}^{s}\mathbf{R}\mathbf{D}^{l_i}),
\end{align}
so that each occurrence of $\mathbf{R}$ is followed by $l_i$ copies of $\mathbf{D}$, where $l_i$ can be zero or higher. The total number of matrices $\mathbf{D}$ or $\mathbf{R}$ has to add up to $k$, which constraints the number of $\mathbf{D}$ that can occur:
\begin{align}
	\sum\limits_{i=1}^{s} l_i=k-s\quad \Rightarrow \quad l_s=k-s-\sum\limits_{i=1}^{s-1}l_i,
	\label{eqn:lindex}
\end{align}
thus summing over all terms at fixed order $k$ involves only $s-1$ independent indices. The total number of terms of order $k$ is:
\begin{align}
	\pmqty{k \\ s}=\frac{1}{s!}k(k-1)\dots (k-s+1),
\end{align}
where the factor $k$ accounts for the $k$ locations in the sequence where the ``first'' occurrence of $\mathbf{R}$ can happen, and the factor $s!$ accounts for the redundant overcounting of the choices of the ``first'', ``second'', and the following occurrences of $\mathbf{R}$, neither of which affect the value of the trace. The remaining counting factors $(k-1),(k-2),\dots$ count the terms with different values of the trace. Putting these contributions together, we can write the following expression for $f_s$:
\begin{align}
	f_s=\Tr \sum\limits_{k=1}^{\infty}\sum\limits_{\{l\}} 
	\prod\limits_{i=1}^{s} \eta^{-2(1+l_i)}\left(\mathbf{R} \mathbf{D}^{l_i}\right)\frac{k}{s!}\frac{(-1)^k}{k},
\end{align}
where the sum over possible sets of $\{l\}$ obeys the constraint of Eqn.~\ref{eqn:lindex}. The counting factor $k$ in the numerator cancels with the factor $k$ from the series expansion of the logarithm. We can now exchange the order of summation in $k$ and $l_i$: instead of doing a constrained sum in all $l_i$ that add up to the same $k$, we treat them on the same level.

Intuitively, the trace can be thought of as a ring that consists of a sequence of $\mathbf{D},\mathbf{R}$ elements. A ring does not have a beginning or the end, hence the factor $k$ in the number of rings with identical value of the trace. The subsequent matrices $\mathbf{R}$ are separated by $l_i$ matrices $\mathbf{D}$. Instead of counting all possible rings of the same length $k$, we instead perform independent sums over the length of all separators $l_i$, and the corresponding rings span all possible lengths, similar to enumeration approaches in heterogeneous self-assembly \cite{murugan2015undesired, klishin2021topological}. We thus rewrite the sum as follows:
\begin{align}
	(-1)^k=&(-1)^s \prod\limits_{i=1}^{s} (-1)^{l_i}\\
	f_s=& \frac{(-1)^s}{s!}\Tr \sum\limits_{\{l\}} \prod\limits_{i=1}^{s} \eta^{-2(1+l_i)} \mathbf{R} (-\mathbf{D})^{l_i}\nonumber\\
	=&\frac{(-1)^s}{s!}\Tr \prod\limits_{i=1}^s \sum\limits_{l_i=0}^{\infty} \eta^{-2(1+l_i)} \mathbf{R} (-\mathbf{D})^{l_i}\nonumber\\
	=&\frac{(-1)^s}{s!}\Tr \left(\left[\eta^{-2} \mathbf{R}(\mathbf{I}+\eta^{-2}\mathbf{D})^{-1}\right]^s\right),
\end{align}
where between the second and third lines we exchanged the order of product and sum. The resulting sum in powers of $(-\mathbf{D})$ is an alternating sign geometric series which converges for large $\eta$, and the resulting inversion of the diagonal matrix $(\mathbf{I}+\eta^{-2}\mathbf{D})$ is valid for any value of $\eta$ because $\mathbf{D}$ is positive semi-definite.

We can thus write the Hamiltonian exactly as follows:
\begin{align}
	\mathcal{H}(\gamma)=&E_b-\Tr\ln(\mathbf{I}+\eta^{-2}\mathbf{D}) \nonumber\\
	+&\sum\limits_{s=1}^{\infty} \frac{(-1)^s}{s!}\Tr\left(\left[\eta^{-2}\mathbf{R} (\mathbf{I}+\eta^{-2}\mathbf{D})^{-1}\right]^s\right),
	\label{eqn:E_resum}
\end{align}
which is a surprisingly concise form in terms of the separated diagonal and off-diagonal terms. If we were to rewrite the matrix expression in index notation, the $s=0$ term would have a single sum over all sensors, and each further term $s$ have a sum over $s$-sensor interactions. Note that since $\mathbf{R}$ is non-diagonal and $(\mathbf{I}+\mathbf{D})^{-1}$ is diagonal, the $s=1$ term is a trace of a non-diagonal matrix and thus it always vanishes. Only the terms for higher $s>1$ have nonzero values. Since the matrix $\mathbf{R}$ does not have a general sign-definite property, the resulting series does not approximate energy either from above or from below.

\subsection{Energy landscapes}
Here we rewrite the matrix expression of Eqn.~\ref{eqn:E_resum} in index notation to highlight the contributions of 1-sensor and 2-sensor terms. First note the following index representation based on Eqn.~\ref{eqn:GGT}:
\begin{align}
	(\mathbf{D}+\mathbf{R})_{ij}=(\mathbf{G}_\gamma \mathbf{G}_\gamma^T)_{ij}=\mathbf{g}_i\cdot \mathbf{g}_j,
\end{align}
where we highlight the nature of the terms as dot products of sensing vectors $\mathbf{g}_i$. The diagonal terms $\mathbf{D}$ correspond to the dot product of each $\mathbf{g}_i$ with \emph{itself}, while the off-diagonal terms $\mathbf{R}$ correspond to the dot products with \emph{different} vectors.

We now need to convert the understanding of the $\mathbf{g}_i$ vectors into the Ising-like Hamiltonian of shape in Eqn.~\ref{eqn:Hgamma}. We ignore the term $E_b$ since it does not depend on the noise or the choice of the sensors. We can also drop the $s=1$ term of the sum since it vanishes because of our matrix decomposition choices. We then truncate the sum in $s$ to only include the term $s=2$, resulting in the following expression for the \emph{2-point energy}:
\begin{align}
	\mathcal{H}_{2pt}(\gamma)\equiv& -\Tr\ln(\mathbf{I}+\mathbf{D}/\eta^2)\nonumber\\
	+&\frac{1}{2}\Tr(\left[\eta^{-2}\mathbf{R} (\mathbf{I}+\eta^{-2}\mathbf{D})^{-1}\right]^2),
	\label{eqn:H2pt}
\end{align}
which we need to transform into the index notation.

The first term has a trace of the log of a diagonal matrix, and thus equals to the sum of the logs of the individual entries:
\begin{align}
	-\Tr\ln(\mathbf{I}+\eta^{-2}\mathbf{D})=&-\sum\limits_{i\in \gamma} \ln(1+\mathbf{D}_i/\eta^2)&\nonumber\\
	=&-\sum\limits_{i\in \gamma} \ln(1+\mathbf{g}_i\cdot \mathbf{g}_i/\eta^2),
\end{align}
which is a sum of 1-sensor terms.

The second term can be rewritten as follows:
\begin{align}
	&\frac{1}{2}\Tr(\left[\eta^{-2}\mathbf{R}(\mathbf{I}+\eta^{-2}\mathbf{D})^{-1}\right]^2)\nonumber \\
	=&\frac{1}{2}\sum\limits_{ij\in \gamma} \frac{\mathbf{R}_{ij}\mathbf{R}_{ji}/\eta^4}{(1+\mathbf{D}_i/\eta^2)(1+\mathbf{D}_j/\eta^2)} \nonumber \\
	=&\frac{1}{2}\sum\limits_{i\neq j \in \gamma} \frac{(\mathbf{g}_i\cdot \mathbf{g}_j)^2/\eta^4}{(1+\mathbf{g}_i\cdot\mathbf{g}_i/\eta^2) (1+\mathbf{g}_j\cdot\mathbf{g}_j/\eta^2)},
\end{align}
where we used the fact that $\mathbf{R}$ is non-diagonal to restrict the sum to only run over the sensor pairs with indices $i\neq j$. For both the 1-sensor and the 2-sensor terms, the sums run only over the sensors in the chosen set $\gamma$. We can thus extend the computation of the terms to the whole landscape that can be analyzed to pick the sensors that optimize the energy. The resulting landscape has the following form for all $ij$:
\begin{align}
	\mathcal{H}_{2pt}(\gamma)=&\sum\limits_{i\in \gamma} h_i + \sum\limits_{i\neq j \in \gamma} J_{ij}\label{eqn:Hgamma_resum}\\
	h_i\equiv& -\ln(1+\mathbf{g}_i\cdot \mathbf{g}_i/\eta^2)\leq 0\label{eqn:hi_resum}\\
	J_{ij}\equiv& \frac{1}{2} \frac{(\mathbf{g}_i\cdot \mathbf{g}_j)^2/\eta^4}{(1+\mathbf{g}_i\cdot\mathbf{g}_i/\eta^2) (1+\mathbf{g}_j\cdot\mathbf{g}_j/\eta^2)} \geq 0,\label{eqn:Jij_resum}
\end{align}
which is valid for any $\eta$.

What are the limits of this energy approximation? The answer to this question is intimately tied to the sensor placement algorithm. Generically, we expect the approximation to work while $\mathbf{g}_i\cdot \mathbf{g}_j\ll \mathbf{g}_i\cdot\mathbf{g}_i$, i.e. the correlation between the sensing vectors is small compared with their magnitude. For many systems it should be possible to choose sensors $i,j$ so that the terms $J_{ij}$ are small compared to the terms $h_i$. However, the \emph{number} of crosstalk terms $J_{ij}$ for $p$ sensors grows as $p^2$ with sensor number $p$. While the individual terms might be small, with increasing number of desired sensors both the number of terms grows, and the algorithm runs out of low crosstalk locations. Due to the combination of these reasons, we expect the approximation to inevitably break down for large number of sensors $p$. How large that number is would depend on the properties of the training data and the placement domain, and thus would need to be established in numerical experiments.

We note that the higher-order terms would have the shape similar to $J_{ij}$ in Eqn.~\ref{eqn:Jij_resum}, with a large number of indices. Due to the construction of $\mathbf{R}$ as a non-diagonal matrix, the terms where \emph{adjacent} indices are identical would correspond to the diagonal of $\mathbf{R}$ and thus vanish. However, the indices can repeat in non-adjacent positions, e.g. at fourth order in $J^{(4)}_{ijij}\neq 0$. The 2-point expression Eqn.~\ref{eqn:Hgamma_resum} is thus not exact even for placement of 2 sensors, but is expected to be a good approximation.


\section{Prior selection}
\begin{figure}
	\begin{center}
		\includegraphics[width=\columnwidth]{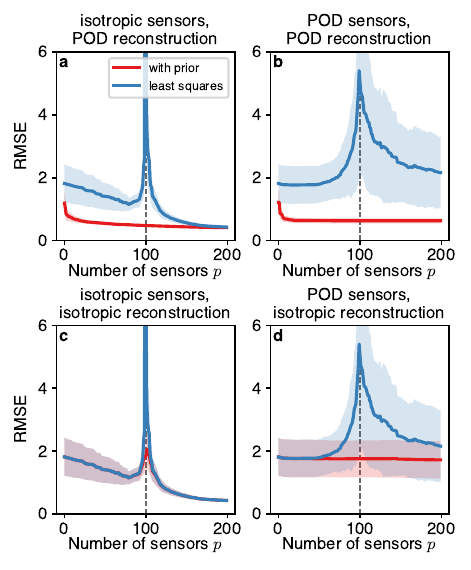}
	\end{center}
	\caption{RMSE benchmark to select the combination of priors. Columns correspond to the choice of prior for sensor placement at noise level \chgtwo{$\eta=1.0^\circ C$}, rows correspond to the choice of prior for reconstruction. Red curves correspond to the reconstruction with the corresponding prior, while blue curves correspond to the least-squares reconstruction. The solid line and shaded region indicating average $\pm$ 1 standard deviation across the test set.}
	\label{fig:prior}
\end{figure}

The reconstruction formula (Eqn.~9 of main text) and the sensor energy landscape both depend on the choice of the prior variances $\mathbf{S}$. We consider the prior to be Gaussian, but it can have different patterns of variances along each dimension: either it is isotropic $\mathbf{S}=\sigma_{prior} \mathbf{I}$ (we choose $\sigma_{prior}=10^3$), or it follows the singular values \chgtwo{(POD)} of the dataset \chgtwo{$\mathbf{S}=\mathbf{\Sigma}_r/\sqrt{N-1}$ \cite{kakasenko2025bridging}}. Since there are two choices of prior for two different operations, we need to consider four possible combinations.

Along with the prior-regularized reconstruction we consider a direct reconstruction of the latent state vector $\hat{\mathbf{a}}$ that solves a linear least square problem:
\begin{align}
	\hat{\mathbf{a}}=\arg \min\limits_{\mathbf{a}} \norm{\mathbf{y}-\mathbf{\Theta}\mathbf{a}}\chgtwo{=\mathbf{\Theta}^\dagger \mathbf{y}},
\end{align}
which is well-defined for any number of sensors $p$.

We show the benchmark comparison in Fig.~\ref{fig:prior}. The least squares reconstruction is never better than regularized reconstruction \chgtwo{and demonstrates the sharp double descent behavior at $p\approx r=100$}.
\chgtwo{The regularized reconstruction curves show strikingly different curves. When an isotropic prior is used for both sensor placement and reconstruction, the peak in RMSE is slightly suppressed but not fully removed (panel c). For the other three scenarios, the RMSE curve is monotonic and has no peak. Placing sensors with a POD prior computes the sensing vector dot products $\mathbf{g}_i\cdot \mathbf{g}_j$ with non-uniform weights, and effectively reduces the dimension in which the sensing vectors are selected (right column of panels). As a result, the selected sensors are more redundant with each other. However, using the POD prior for reconstruction significantly reduces the RMSE while removing the double descent peak (top row of panels). As a result, using an isotropic prior for sensor placement results in a wide variety of non-redundant sensors selected, while the POD regularization improves reconstruction (panel a).}

Following the conclusions of this benchmark test, we pick sensors against an isotropic prior, yet reconstruct states with the POD prior. This combination of methods is used for all results reported in the main text and Figs.~\ref{fig:faces}-\ref{fig:cylinder}.





\section{Sensor method comparison}

\begin{figure*}
	\begin{center}
		\includegraphics[width=\textwidth]{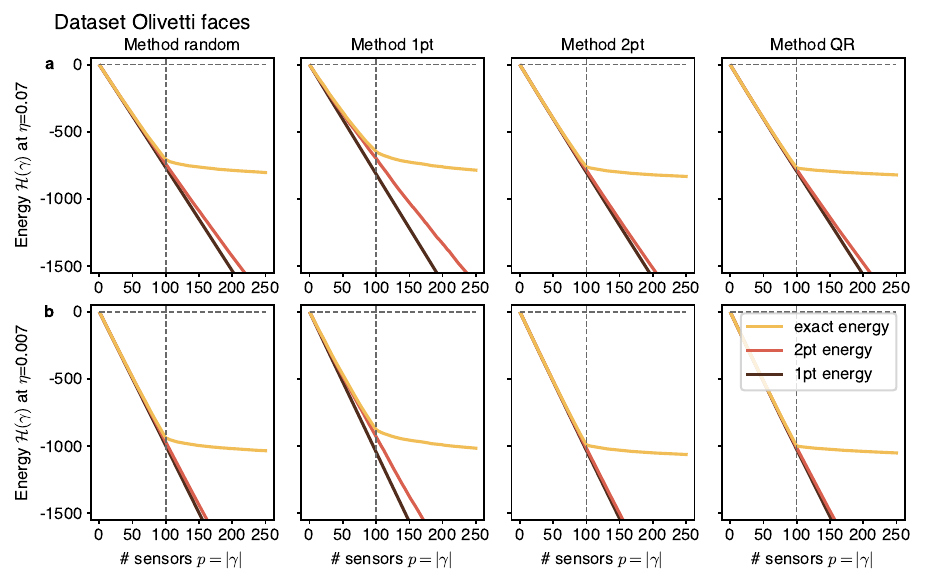}
	\end{center}
	\caption{\chgtwo{Comparison of sensor placement methods for the Olivetti faces dataset as measured by different energy expressions. The rows (a) and (b) use the energy landscape at different values of noise $\eta$ ($0.5 \sigma_{scale}$ and $0.05\sigma_{scale}$, respectively), the columns correspond to four different sensor placement methods. The curves correspond to three energy formulae. Exact energy is evaluated by directly computing the determinant in Eqn.~\ref{eqn:Hgamma}, ignoring the constant term $E_b$. 2-point energy is evaluated with Eqn.~\ref{eqn:Hgamma_resum}. 1-point energy is evaluated by taking only the first term in Eqn.~\ref{eqn:Hgamma_resum}. The horizontal dashed line indicates $\mathcal{H}=0$, the vertical line indicates the reconstruction rank $p=r$.}}
	\label{fig:methods_faces}
\end{figure*}

\begin{figure*}
	\begin{center}
		\includegraphics[width=\textwidth]{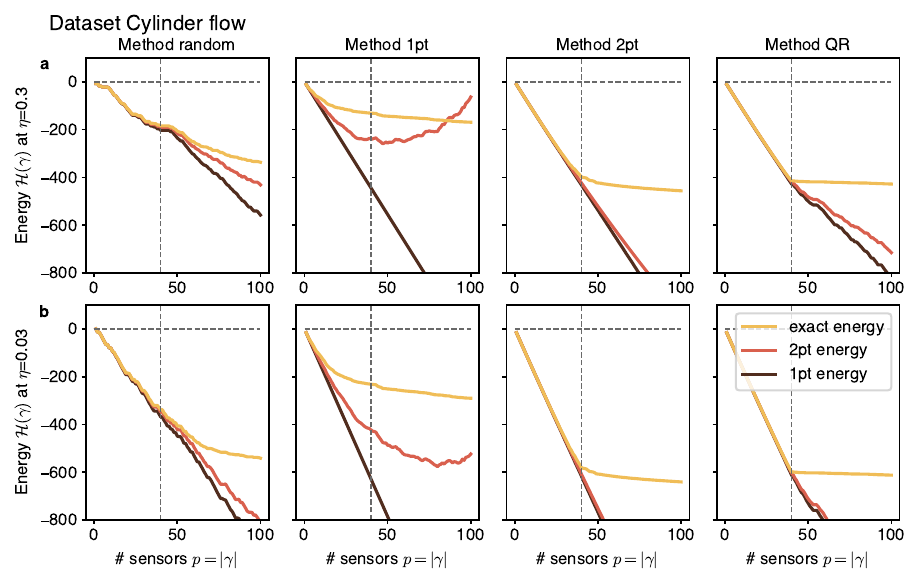}
	\end{center}
	\caption{\chgtwo{Comparison of sensor placement methods for the cylinder flow dataset as measured by different energy expressions. The rows (a) and (b) use the energy landscape at different values of noise $\eta$ ($0.5 \sigma_{scale}$ and $0.05\sigma_{scale}$, respectively), the columns correspond to four different sensor placement methods. The curves correspond to three energy formulae. Exact energy is evaluated by directly computing the determinant in Eqn.~\ref{eqn:Hgamma}, ignoring the constant term $E_b$. 2-point energy is evaluated with Eqn.~\ref{eqn:Hgamma_resum}. 1-point energy is evaluated by taking only the first term in Eqn.~\ref{eqn:Hgamma_resum}. The horizontal dashed line indicates $\mathcal{H}=0$, the vertical line indicates the reconstruction rank $p=r$.}}
	\label{fig:methods_cylinder}
\end{figure*}

\chgtwo{Figs.~\ref{fig:methods_faces}-\ref{fig:methods_faces} show the comparison of sensor placement methods and respective energies for other two empirical datasets in the same format as Fig.~2 of the main text.}

\section{Sensor placement for Olivetti faces}
\begin{figure*}
	\begin{center}
		\includegraphics[width=\textwidth]{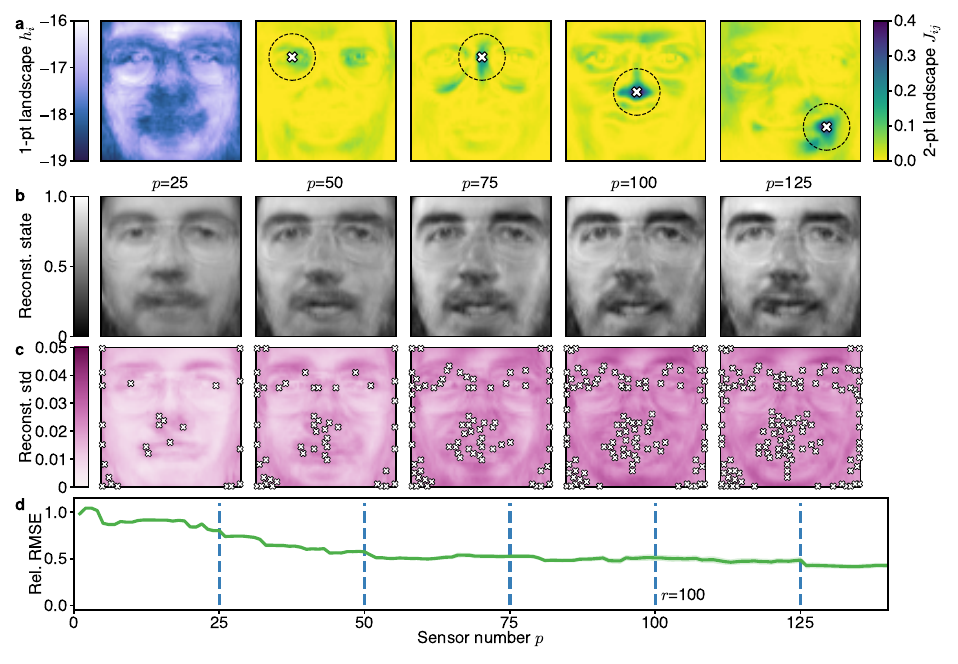}
	\end{center}
	\caption{\chgtwo{Reconstruction of Olivetti faces with data-driven sensor placement with $r=100$ modes and noise $\eta=0.2\sigma_{scale}$. (a) The 1-point landscape $h_i$ (left panel) and the 2-point landscape $J_{ij}$ conditioned on the sensor locations marked with white crosses and highlighted with dashed circles (four right panels). (b) Maximal likelihood reconstruction. (c) Reconstruction uncertainty heatmap with white crosses marking the sensors selected by the 2-point algorithm. (d) Relative root mean square error (RMSE) of state reconstruction, with solid line and shaded region indicating average $\pm$ 1 standard deviation across the test set. The vertical dashed lines indicates the sensor counts for the reconstructions shown in (b)-(c).}}
	\label{fig:faces}
\end{figure*}

Fig.~\ref{fig:faces} shows the sensor placement and reconstruction for the Olivetti faces dataset. \chgtwo{The 1-point landscape $h_i$ has many localized basins for placement, aligning with the areas of high variability in human faces: eyes, nose, lips, as well as corners of the image (panel a). The 2-point landscape $J_{ij}$ shows both the local repulsion of sensors from nearby locations, and the facial symmetry features such as similarity between the two eyes (panel a). As the number of sensors increases, the reconstructed face gets progressively sharper features (panel b). The sensor placement largely follows the many local basins of the 1-point landscape (panel c). Since the basins are not very deep and sensor repulsion precludes placing many sensors close by, the 2-point algorithm effectively distributes the sensors throughout all highly-variable features across the domain and gradually increases the placement density with higher sensor budgets. The uncertainty magnitude \emph{increases} (darker color) with higher sensor counts, while the RMSE curve shows small local fluctuations along with the overall nearly monotonic decrease. This latter property suggests that while our algorithm identifies the principal facial features, the Olivetti faces are not well-described with 100 linear modes, and either a higher mode count or a nonlinear encoding are required.}


\section{Sensor placement for cylinder flow}
\begin{figure*}
	\begin{center}
		\includegraphics[width=\textwidth]{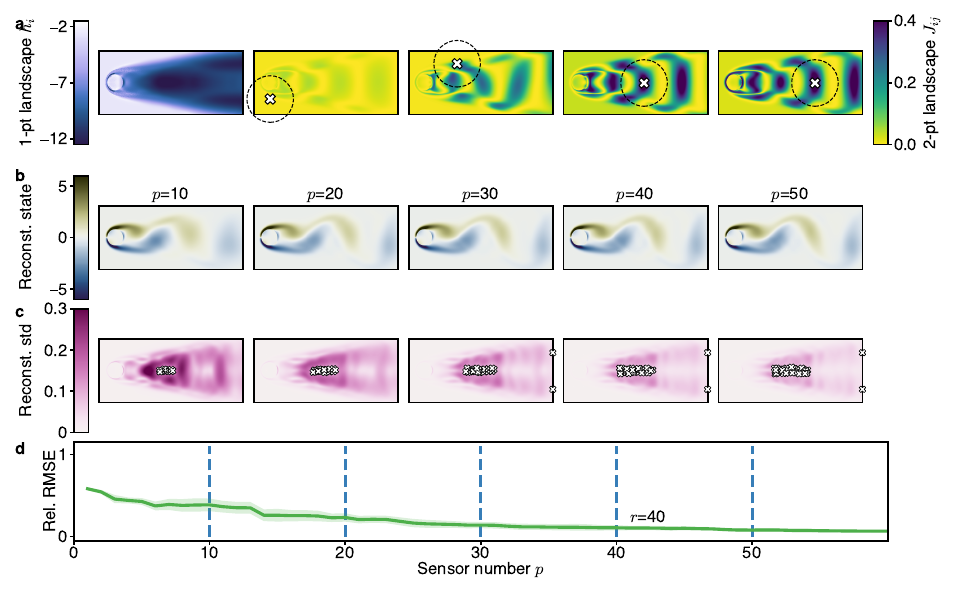}
	\end{center}
	\caption{\chgtwo{Reconstruction of cylinder flow with data-driven sensor placement with $r=40$ modes and noise $\eta=0.2\sigma_{scale}$. (a) The 1-point landscape $h_i$ (left panel) and the 2-point landscape $J_{ij}$ conditioned on the sensor locations marked with white crosses and highlighted with dashed circles (four right panels). (b) Maximal likelihood reconstruction. (c) Reconstruction uncertainty heatmap with white crosses marking the sensors selected by the 2-point algorithm. (d) Relative root mean square error (RMSE) of state reconstruction, with solid line and shaded region indicating average $\pm$ 1 standard deviation across the test set. The vertical dashed lines indicates the sensor counts for the reconstructions shown in (b)-(c).}}
	\label{fig:cylinder}
\end{figure*}

Fig.~\ref{fig:cylinder} shows the sensor placement and reconstruction for the cylinder flow dataset. \chgtwo{The 1-point landscape shows a deep basin along the symmetry axis in the wake behind the cylinder, as well as two more shallow basins towards the right side of the domain symmetrically offset from the axis (panel a). The 2-point landscape shows that locations outside of the wake have practically no correlation with those inside the wake (panel a). Within the wake, the correlation structure has a distinct wave-like pattern corresponding to the spatial wavelength of the vortex shedding (panel a). The reconstructions show an error imperceptible by eye already at $p=10$ sensors (panel b). Unlike SST and Olivetti faces datasets, the cylinder flow dataset shows a tight clustering of sensors that gradually gets longer and wider, with only two peripheral sensors appearing for $20<p<30$ (panel c). The RMSE curve decays monotonically towards a very low value without significant fluctuations.}

\chgtwo{The sensor placement pattern can be explained by the quantitative features of the landscape. The central basin of 1-point energy is both deep and wide, reaching the lowest value of $h_{center~min}\approx -12.97$. The 2-point interactions of the sensors have a narrow band structure corresponding to the wake wavelength, but in magnitude peak at about $J_{ij}\approx 0.4$. It is thus advantageous to place multiple sensors very close to each other up to a point, until the crosstalk of a new sensor with previously selected neighbors overwhelms the advantage of the basin. As a result, as the sensor budget grows, sensors are placed in a band that gradually gets both longer and wider (along and orthogonally to the symmetry axis). Fig.~\ref{fig:cylinder} also shows that the sensor placement method that accounts for crosstalk (2-point or QR) is significantly better than those that do not (random and 1-point) in terms of reaching lower energy at the same sensor budget.}

\chgtwo{Beyond the central basin, the landscape also shows two symmetric secondary basins offset from the central axis on the far-right end of the domain. The minimal value of the 1-point term $h_i$ is reached in the last column and is equal to $h_{side~min}\approx -12.30$. Thus, it is advantageous for the algorithm to place sensors within the central basin until it is exhausted, i.e., when the crosstalk with previously placed sensors cancels out the difference between the central and side basins $J>(h_{side~min}-h_{center~min})\approx 0.7$. In the regime presented in Fig.~S6 this corresponds to sensors number 25 and 28 so they appear in the third visualized frame at $p=30$.}


\section{Exhaustive sensor enumeration}
\begin{figure*}
	\begin{center}
		\includegraphics[width=\textwidth]{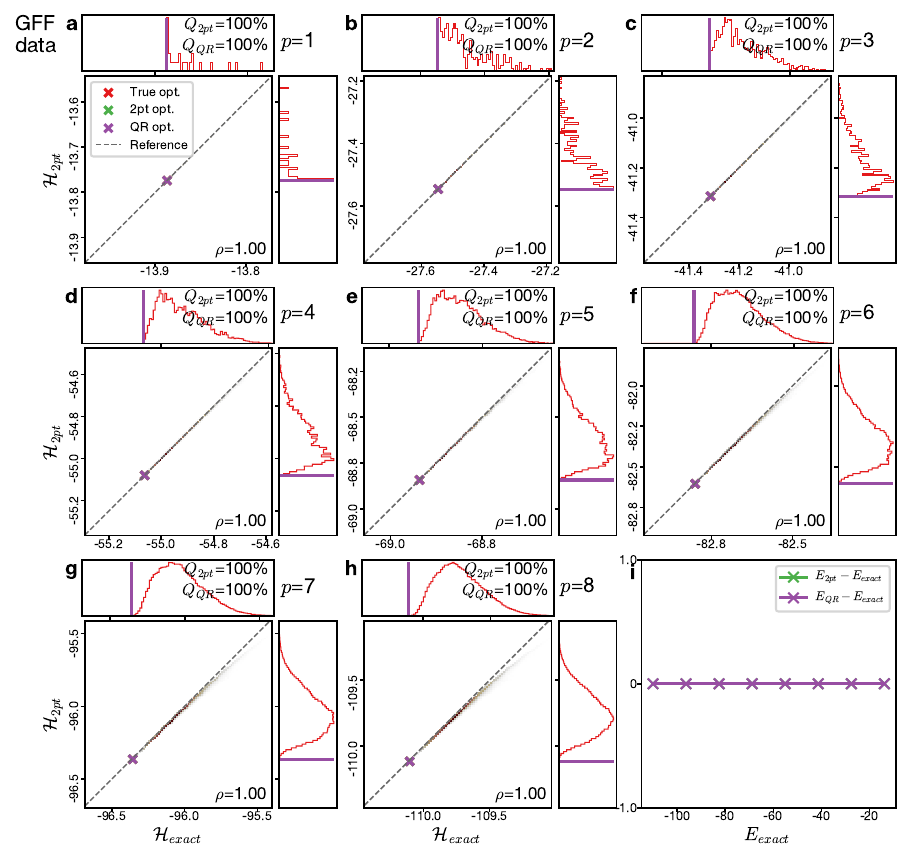}
	\end{center}
	\caption{Comparison of sensor configurations between brute force search, 2-point algorithm, and QR methods for the GFF dataset. (a-h) Two-dimensional histograms of the joint distribution in the exact energy \eqref{eqn:Hgamma} and 2-point energy $\eqref{eqn:H2pt}$ for all configurations with sensor number $p\in[1,8]$. The value of $\rho$ corresponds to the Spearman correlation between the two energies. The right and top panels show the marginal histogram in only one of the variables. The crosses indicate the sensor configurations found by each of the three methods. \chgtwo{All three crosses overlap in each panel since both the 2-point and the QR method find the true best sensor set in this case.} (i) The discrepancy in energy between the absolute and approximate minima found by the two methods. Lower exact energy corresponds to more sensors.}
	\label{fig:hist_GFF}
\end{figure*}

\begin{figure*}
	\begin{center}
		\includegraphics[width=\textwidth]{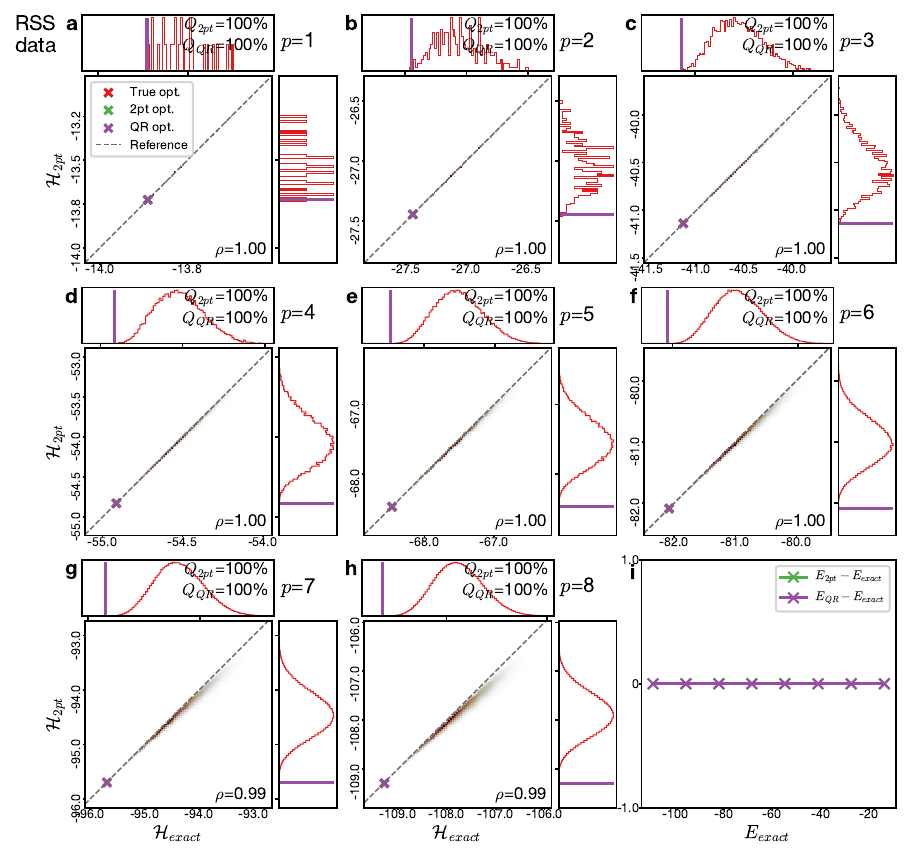}
	\end{center}
	\caption{Comparison of sensor configurations between brute force search, 2-point algorithm, and QR methods for the RSS dataset. (a-h) Two-dimensional histograms of the joint distribution in the exact energy \eqref{eqn:Hgamma} and 2-point energy $\eqref{eqn:H2pt}$ for all configurations with sensor number $p\in[1,8]$. The value of $\rho$ corresponds to the Spearman correlation between the two energies. The right and top panels show the marginal histogram in only one of the variables. The crosses indicate the sensor configurations found by each of the three methods. \chgtwo{All three crosses overlap in each panel since both the 2-point and the QR method find the true best sensor set in this case.} (i) The discrepancy in energy between the absolute and approximate minima found by the two methods. Lower exact energy corresponds to more sensors.}
	\label{fig:hist_RSS}
\end{figure*}
\label{sec:exhaustive}

Figs.~2,4 of main text establish the near-equivalence of the 2-point algorithm and the QR algorithm for sensor placement, but both algorithms are approximate optimizers. For $p$ sensors in $n$-dimensional state space there exist $\pmqty{n \\ p}$ possible configurations that can be enumerated directly for small $n,p$. In Figs.~\ref{fig:hist_GFF},\ref{fig:hist_RSS} we enumerate all sensor configurations for the GFF and RSS datasets, respectively, which both have $n=25$ and $p\in[1,8]$.

We assess the degree of optimality of the two algorithms with three metrics. First, we compute the Spearman rank correlation $\rho$ between the exact and approximate 2-point energies, which indicates to what degree minimizing the 2-point energy (moving down on the two-dimensional histograms) is equivalent to minimizing the true energy (moving left on the histograms). The Spearman correlation stays at the level of $\rho>0.99$ to two significant figures, showing that the exact energy is well approximated by the 2-point energy \emph{for all sensor configurations}.

Second, we compute its \emph{efficiency} $Q$, equal to the fraction of the sensor configurations with exact energy equal or higher than the configuration found by a particular method. For the best solution found by brute force search, efficiency is by definition $Q=100\%$. Across both datasets, efficiency of the 2-point and QR methods reaches that level, indicating that both methods find the exact optimal sensor configuration.

Third, we compute the energy difference between the true minimum and the approximations (Figs.~\ref{fig:hist_GFF}i,\ref{fig:hist_RSS}i) that remains at the level of floating point number precision.

An important limitation of this comparison is the artificially small size of the GFF and RSS datasets. While it is easy to generate larger datasets with the same software, they would not be amenable to brute force enumeration. Within the empirical datasets such as the Sea Surface Temperature, it is easy to place sensors with low crosstalk (Fig.~1 of main text) since crosstalk typically falls off with spatial distance between sensor locations. For small datasets, there are no locations with large distance between them, and thus it is impossible to achieve low crosstalk. We are looking forward to future methods for comparing the exact and approximately optimal sensor configurations.

\bibliography{sensors}